\documentclass[aps,prd,twoside,twocolumn,nofootinbib,showpacs,floatfix]{revtex4-1}
\usepackage{amssymb}
\usepackage{amsmath,amssymb}
\usepackage{graphicx,bm}
\usepackage{slashed}
\usepackage{times}
\usepackage{epstopdf}
\usepackage{ulem} 
\usepackage[usenames]{color}
\usepackage{float}
\usepackage{subfigure}
\usepackage{subfigure}
\usepackage{rotating}
\usepackage{color}
\usepackage{multirow}
\usepackage{dcolumn}
\usepackage{overpic}
\usepackage{booktabs}
\usepackage{appendix}
\usepackage{makecell}
\usepackage{textcomp}
\usepackage{diagbox}

\renewcommand\sout{\bgroup \color{red} \ULdepth=-.5ex \ULset}

\usepackage[colorlinks, citecolor=blue,anchorcolor=red,menucolor=red,linkcolor=blue,filecolor=red,runcolor=red,urlcolor=blue,frenchlinks=red]{hyperref}
\begin{document}
\title{Classifying the hidden-charm pentaquarks via a flavor mixing scheme}
\author{Kan Chen}\email{chenk10@nwu.edu.cn}
\affiliation{School of Physics, Northwest University, Xi'an 710127, China}
\affiliation{Shaanxi Key Laboratory for theoretical Physics Frontiers, Xi'an 710127, China}
\affiliation{Institute of Modern Physics, Northwest University, Xi'an 710127, China}
\affiliation{Peng Huanwu Center for Fundamental Theory, Xi'an 710127, China}
\author{Bo Wang}\email{wangbo@hbu.edu.cn}
\affiliation{College of Physics Science \& Technology, Hebei University, Baoding 071002, China}
\affiliation{Hebei Key Laboratory of High-precision Computation and Application of Quantum Field Theory, Baoding, 071002, China}
\affiliation{Hebei Research Center of the Basic Discipline for Computational Physics, Baoding, 071002, China}

\begin{abstract}
In this work, we propose a scheme to classify the molecular states consisting of ground single-charm baryons ($\Lambda_c$, $\Xi_c$, $\Sigma_c^{(*)}$, $\Xi_c^{\prime(*)}$, $\Omega_c^{(*)}$) and $\bar{D}^{(*)}/\bar{D}_s^{(*)}$ mesons. Within this framework, all considered baryon-meson systems are categorized according to the flavor components of their light degrees of freedom. We briefly illustrate how this classification scheme can consistently explain the experimentally observed $P_c$ and $P_{cs}$ states. This framework also predicts the existences of single-strange and double-strange hidden-charm bound states. The attractive interactions of these states arise from channel mixing between $\Sigma_c^{(*)}\bar{D}_s^{(*)}$ and $\Xi_c^{\prime(*)}\bar{D}^{(*)}$ for single-strange systems, and mixing between $\Xi_c^{\prime(*)}\bar{D}_s^{(*)}$ and $\Omega_c^{(*)}\bar{D}^{(*)}$ for double-strange systems, respectively. Using parameters fitted from the measured $P_c$ and $P_{cs}$ states, we systematically present the predicted mass spectra for these single- and double-strange hidden-charm bound states.
\end{abstract}

\maketitle

\vspace{2cm}

\section{Introduction}\label{sec1}
Since 2015, numerous hidden-charm pentaquark states have been observed in experiments, including the $P_c(4312)$, $P_c(4380)$, ($P_c(4440)$, $P_c(4457)$) \cite{LHCb:2015yax,LHCb:2016ztz,LHCb:2019kea} pentaquark candidates, and the $P_{cs}(4338)$ \cite{LHCb:2022ogu}, $P_{cs}(4459)$ \cite{LHCb:2020jpq,Belle:2025pey} single-strange pentaquark candidates. Various theoretical models have been put forward to interpret the underlying nature of these two categories of pentaquark candidates (see Ref. \cite{Chen:2016qju,Lebed:2016hpi,Esposito:2016noz,Guo:2017jvc,Ali:2017jda,Liu:2019zoy,Chen:2022asf,Meng:2022ozq,Wang:2025dur} for relevant reviews). The masses of these pentaquark candidates lie very close to the thresholds of $\Sigma_c\bar{D}$, $\Sigma_c^*\bar{D}$, $\Sigma_c\bar{D}^*$, $\Xi_c\bar{D}$, and $\Xi_c\bar{D}^*$, respectively. Accordingly, they serve as good molecular candidates consisting of an anticharmed meson and a single-charm baryon.

With a growing number of hidden-charm molecular states anticipated in upcoming experimental measurements, establishing a systematic and efficient classification framework for molecular pentaquark candidates becomes an essential theoretical task.

Within the conventional quark model \cite{Gell-Mann:1964ewy,Zweig:1964ruk,Zweig:1964jf}, a meson is composed of a quark-antiquark pair, the direct product of the flavor $\bm{3}$ and $\bar{\bm{3}}$ representations yields a singlet and an octet. A baryon is composed of three quarks, the direct product of three flavor $\bm{3}$ representations gives rise to a singlet, two octets, and a decuplet. To date, almost all light-flavor mesons and baryons can be successfully classified within this well-established framework.

Small binding energies are the most notable feature of molecular states, and they usually arise from light meson exchanges. For hidden-charm molecular candidates within such a meson exchange picture, the exchange of heavy $c\bar{c}$ meson is suppressed by a factor of $1/m_{c\bar{c}}^2$ ($m_{c\bar{c}}$ denotes the mass of the exchanged heavy $c\bar{c}$ meson) and thus is generally omitted from the full effective potential. Consequently, light $q\bar{q}$ meson exchanges ($q=u$, $d$, $s$) dominate the effective potential.

When light mesons are exchanged between baryons and mesons, the light quark constituents inside mesons and baryons couple to these exchanged light mesons. Since SU(3) flavor symmetry enables us to classify the light quark components of baryons and mesons, the same symmetry also enable us to categorize the flavors of light mesons that can be exchanged within baryon-meson systems.

If the above discussion captures some of the properties of the interactions between hidden-charm molecular pentaquarks, then, similar to the classification of conventional mesons and baryons via SU(3) flavor symmetry, SU(3) flavor symmetry may also serve as the key to categorizing hidden-charm molecular states.

In Ref. \cite{Chen:2022wkh}, to investigate the relations between the $P_c$ and $P_{cs}$ states, we proposed a contact potential possessing SU(3) flavor symmetry and SU(2) spin symmetry to describe the interactions of the $P_c$ and $P_{cs}$ states. We demonstrated that since the $\Sigma_c$ and $\Xi_c^\prime$ belong to the same SU(3) sextet, while the $\Xi_c$ baryon belongs to the SU(3) triplet. For this reason, instead of the $\Xi_c\bar{D}^{(*)}$ system, the spectra of $\Sigma^{(*)}_c\bar{D}^{(*)}$ and $\Xi_c^{\prime(*)}\bar{D}^{(*)}$ manifest similar behaviors. Consequently, SU(3) flavor symmetry indicates that the observed $P_c$ states can not be related to the observed $P_{cs}$ states. Thus, our previous analysis left us with the question of how to address the relations between the observed $P_c$ and $P_{cs}$ states.

With the scheme introduced in this work, we will show that the observed $P_c$ states lie in the SU(3) $|\bm{8}^\prime,1,\frac{1}{2}\rangle$ doublet (we clarify this notation in Sec. \ref{sec22}), while the observed $P_{cs}$ states belong to the broken $|\bm{1},0,0\rangle$ singlet. Besides, we present two other groups of single-strange hidden-charm molecular states consisting of $\Xi_c^{\prime(*)}\bar{D}^{(*)}$ and $(\Xi_c^{\prime(*)}\bar{D}^{(*)},\,\Sigma_c^{(*)}\bar{D}_s^{(*)})$ components, which belong to the $|\bm{8}^{\prime},0,0\rangle$ singlet and broken $|\bm{8}^\prime,0,1\rangle$ triplet, respectively.

Our classification scheme also leads to the existences of double-strange hidden-charm molecular states, which consist of ($\Xi_c^{\prime(*)}\bar{D}_s^{(*)}$, $\Omega_c^{(*)}\bar{D}^{(*)}$) components and belong to the broken $|\bm{8}^\prime,-1,\frac{1}{2}\rangle$ doublet. Theoretically, the double-strange hidden-charm pentaquark states have been explored within various frameworks, including the one-boson-exchange model \cite{Wang:2020bjt}, unitarized coupled-channel hidden gauge formalism \cite{Marse-Valera:2022khy,Roca:2024nsi}, off-shell coupled-channel approach \cite{Clymton:2025zer}, QCD sum rule \cite{Azizi:2021pbh,Wang:2026thx,Wang:2025pjt}, quark model \cite{Anisovich:2015zqa,Ortega:2022uyu}, and also a recent study by assuming the $c\bar{c}ssq$ system as a compact configuration \cite{Mutuk:2026fwg}.

This work is organized as follows. In Sec. \ref{sec2}, we introduce our theoretical framework. In Secs. \ref{sec3}-\ref{sec4}, we briefly recap our previous descriptions of the $P_c$ and $P_{cs}$ states and present the classification scheme for hidden-charm molecular pentaquarks. Our discussions and predictions for possible bound states arising from the $\Sigma_c^{(*)}\bar{D}_s^{(*)}$-$\Xi_c^{\prime(*)}\bar{D}^{(*)}$ and $\Xi_c^{\prime(*)}\bar{D}_s^{(*)}$-$\Omega_c^{(*)}\bar{D}^{(*)}$ mixings are given in Sec. \ref{sec5}. Sec. \ref{sec6} is devoted to a summary.

\section{Framework}\label{sec2}
In this section, we present the Lagrangian describing the interactions between single-charm ($\Lambda_c$, $\Xi_c$, $\Sigma_c^{(*)}$, $\Xi_c^{\prime(*)}$, $\Omega_c^{(*)}$) baryons and anticharmed ($\bar{D}^{(*)}$, $\bar{D}_s^{(*)}$) mesons. We then construct the flavor $\otimes$ spin wave functions for these baryon-meson systems.

In this work, the flavor wave functions of the considered baryon-meson systems are built upon two symmetry assumptions, namely exact SU(3) symmetry (SU(3) basis) and broken SU(3) symmetry (isospin basis). By comparing the explicit expressions of baryon-meson flavor wave functions within the SU(3) basis and isospin basis, we illustrate how the physical flavor wave functions in the isospin basis descend from the SU(3) flavor wave functions that possess higher symmetry.

\subsection{Effective potential}\label{sec21}
The interactions between single-charm baryons and anticharmed mesons can be derived by considering the effective potentials originating from light meson exchanges. To investigate bound states lying very close to a baryon-meson threshold, we adopt the $S$-wave quark-level Lagrangian proposed in Refs. \cite{Chen:2022wkh,Chen:2024bre}.
\begin{eqnarray}
\mathcal{L}=g_s\bar{q}\mathcal{S}q+g_a\bar{q}\gamma_\mu\gamma^5\mathcal{A}^\mu q.\label{Lagrangians}
\end{eqnarray}
Here, the matrices $\mathcal{S}$ and $\mathcal{A}^\mu$ are the fictitious scalar ($J^P=0^+$) and axial-vector ($J^P=1^+$) fields.

The effective potential of the light quark-quark interactions from the Lagrangian in Eq. (\ref{Lagrangians}) can be written as
\begin{eqnarray}
V_{qq}=\tilde{g}_s\bm{\lambda}_1\cdot\bm{\lambda}_2+\tilde{g}_a\bm{\lambda}_1\cdot\bm{\lambda}_2\bm{\sigma}_1\cdot\bm{\sigma}_2,
\end{eqnarray}
with
\begin{eqnarray}
\bm{\lambda}_1\cdot\bm{\lambda}_2=\lambda_1^8\lambda_2^8+\sum_{i=1}^3\lambda_1^i\lambda_2^i+\sum_{j=4}^7\lambda_1^j\lambda_2^j,
\end{eqnarray}
and
\begin{eqnarray}
\bm{\sigma}_1\cdot\bm{\sigma}_2=\sum_{i=1}^{3}\sigma_1^i\sigma_2^i.
\end{eqnarray}
The operators $\lambda_1^8\lambda_2^8$ ($\lambda_1^8\lambda_2^8\bm{\sigma}_1\cdot\bm{\sigma}_2$), $\sum_{i=1}^3\lambda_1^i\lambda_2^i$ ($\sum_{i=1}^3\lambda_1^i\lambda_2^i\bm{\sigma}_1\cdot\bm{\sigma}_2$), and $\sum_{j=4}^7\lambda_1^j\lambda_2^j$ ($\sum_{j=4}^{7}\lambda_1^j\lambda_2^j\bm{\sigma}_1\cdot\bm{\sigma}_2$) account for the exchanges of isospin singlet, triplet, and two doublets light scalar (axial-vector) fictitious meson fields.

The redefined coupling parameters $\tilde{g}_s$ and $\tilde{g}_a$ are proportional to $\dfrac{g_s^2}{m_{\mathcal{S}}^2}$ and
$\dfrac{g_a^2}{m_{\mathcal{A}}^2}$, respectively, where $m_{\mathcal{S}}$ and $m_{\mathcal{A}}$ denote the masses of the fictitious scalar and axial-vector mesons. The magnitudes and signs of $\tilde{g}_s$ and $\tilde{g}_a$ can be determined by the observed $P_c$ states \cite{Chen:2024tuu}. This potential respects SU(3) flavor symmetry and SU(2) spin symmetry for interactions between all baryon-meson systems considered here.

Note that the operators $\sum_{i=1}^3\lambda_1^i\lambda_2^i$ and $\sum_{j=4}^7\lambda_1^j\lambda_2^j$ describe the non-strange light meson exchange and the strange light meson exchange, respectively. To incorporate SU(3) symmetry breaking originating from strange meson exchange, we assume that the coupling parameters $\tilde{g}_s$ for these two operators are different. Explicitly, comparing to the non-strange light meson exchange, the coupling parameter for the strange meson exchange is suppressed by the larger mass of the exchanged light strange meson, i.e., suppressed by the factor
\begin{eqnarray}
g_x=\frac{1}{m_{\text{s}}^2}/\frac{1}{m^2_{\text{nons}}}=\frac{m^2_{\text{nons}}}{m^2_{\text{s}}},
\end{eqnarray}
where $m_{\text{s}}$ and $m_{\text{nons}}$ are the masses of the exchanged strange and non-strange light mesons, respectively. The suppression factor $g_x$ should lie in the $[0,1]$ region, and in the SU(3) limit we have $g_x=1$.
\subsection{Flavor and spin wave functions}\label{sec22}
To investigate the symmetry properties of flavor wave functions for systems consisting of an anticharmed meson and a single-charm baryon, we first construct the baryon-meson wave functions in the SU(3) flavor limit (we refer to the flavor functions constructed under this symmetry assumption as the SU(3) basis), i.e., the light diquark components inside single-charm baryons transform as the $\bar{\mathbf{3}}$ or $\mathbf{6}$ representation,
\begin{eqnarray}
\textbf{3}\otimes\textbf{3}=\bar{\textbf{3}}\oplus\textbf{6}.
\end{eqnarray}
The light quark flavor of an anticharmed meson is $u$, $d$, or $s$ and transforms as the $\mathbf{3}^\prime$ representation. Combining the light diquark components inside a single-charm baryon with light quark inside an anticharmed meson, we can obtain the $\mathbf{1}$, $\mathbf{8}$, $\mathbf{8}^\prime$, and $\mathbf{10}$ representations
\begin{eqnarray}
\bar{\textbf{3}}\otimes\textbf{3}^\prime&=&\textbf{1}\oplus \textbf{8},\\
\textbf{6}\otimes\textbf{3}^\prime&=&\textbf{8}^\prime\oplus\textbf{10}.
\end{eqnarray}
The flavor wave functions of hidden-charm baryon-meson systems within the SU(3) basis can be derived using standard SU(3) Clebsch-Gordan (CG) coefficients \cite{Kaeding:1995vq}.

Next, we turn to physical states, where we must treat the $s$ quark separately from the $u$ and $d$ quarks. For $\Lambda_c$ and $\Sigma_c^{(*)}$ baryons, SU(2) CG coefficients suffice to build their flavor wave functions. For the $\Xi_c$/$\Xi_c^{\prime(*)}$ baryons, we explicitly antisymmetrize/symmetrize the $s$ quark with the light $n$ quark ($n=u$ or $d$) to construct the corresponding flavor wave functions.

The explicit flavor wave functions for the single-charm baryons and anticharmed mesons used within the isospin basis formalism are collected in Table \ref{flavor WF}. The pair $(\bar{D}^{(*)0},\,\bar{D}^{(*)-})$ forms a flavor doublet, whereas $\bar{D}_s^{(*)-}$ corresponds to a flavor singlet. As also summarized in Table \ref{flavor WF}, the two light quarks inside each single-charm baryon carry well-defined symmetry properties. Making use of these tabulated flavor wave functions, we construct the full baryon-meson flavor wave functions using SU(2) CG coefficients (we refer to the wave functions built under this symmetry assumption as the isospin basis).

\begin{table}[htbp]
\setlength\tabcolsep{0.9pt} \caption{The flavor wave functions of
the anticharmed mesons and single-charm baryons used in the isospin basis. \label{flavor WF}}
\begin{tabular}{llc|llc}
\toprule[1pt]
Hadron&$\left|Im_I\right\rangle$&$\phi^{M}_{Im_I}$&Hadron&$\left|Im_I\right\rangle$&$\phi^{M}_{Im_I}$\\
\hline
$\bar{D}^{(*)0}$&$\left|\frac{1}{2}\frac{1}{2}\right\rangle$&$u\bar{c}$&$\bar{D}^{(*)-}$&$\left|\frac{1}{2}-\frac{1}{2}\right\rangle$&$d\bar{c}$\\
$\bar{D}_s^{(*)-}$&$\left|00\right\rangle$&$s\bar{c}$&&&\\
\hline
Hadron&$\left|Im_I\right\rangle$&$\phi^{B}_{Im_I}$&Hadron&$\left|Im_I\right\rangle$&$\phi^{B}_{Im_I}$\\
\hline
$\Lambda_c^+$&$\left|00\right\rangle$&$\frac{1}{\sqrt{2}}\left(du-ud\right)c$&$\Sigma_c^{(*)++}$&$\left|11\right\rangle$&$uuc$\\
$\Sigma_c^{(*)+}$&$\left|10\right\rangle$&$\frac{1}{\sqrt{2}}\left(ud+du\right)c$&$\Sigma_c^{(*)0}$&$\left|1-1\right\rangle$&$ddc$\\
$\Xi_c^+$&$\left|\frac{1}{2}\frac{1}{2}\right\rangle$&$\frac{1}{\sqrt{2}}\left(us-su\right)c$&$\Xi_c^0$&$\left|\frac{1}{2}-\frac{1}{2}\right\rangle$&$\frac{1}{\sqrt{2}}\left(ds-sd\right)c$\\
$\Xi_c^{\prime(*)+}$&$\left|\frac{1}{2}\frac{1}{2}\right\rangle$&$\frac{1}{\sqrt{2}}\left(us+su\right)c$&$\Xi_c^{\prime(*)0}$&$\left|\frac{1}{2}-\frac{1}{2}\right\rangle$&$\frac{1}{\sqrt{2}}\left(ds+sd\right)c$\\
$\Omega_c^{(*)0}$&$\left|00\right\rangle$&$ssc$\\
\bottomrule[1pt]
\end{tabular}
\end{table}

By comparing the explicit expressions of flavor wave functions built within the SU(3) basis and isospin basis, we derive the relations between the flavor wave functions in these two bases. Table \ref{SU3ISO} expresses baryon-meson flavor wave functions from the SU(3) basis as linear combinations of isospin-basis states. We adopt the notation $|\bm{R},Y,I\rangle$ \cite{Kaeding:1995vq}$\,$ (or $|\bm{R},Y,I,I_z\rangle$) for SU(3)-basis flavor wave functions. Here, $Y$ stands for hypercharge defined via $Y=B+S$, where $B$ is baryon number and $S$ denotes strangeness. The symbols $\bm{R}$, $I$, and $I_z$ correspond to the dimension of multiplet, total isospin, and third isospin component, respectively. As illustrated in Table \ref{SU3ISO}, every $|\bm{R},Y,I\rangle$ SU(3) eigenstate reduces to either a single baryon-meson isospin state or a linear superposition of two such isospin-basis states.

\begin{table}[htbp]
\renewcommand\arraystretch{1.5}
\caption{The baryon-meson flavor wave functions constructed in the SU(3) basis represented in terms of the isospin basis.}
\begin{tabular}{c|cccccccccccccc}
\toprule[0.8pt]
$|R,Y,I\rangle$&Wave function\\
\hline
$|\bm{1},0,0\rangle$&$\sqrt{\frac{1}{3}}|\Lambda_c\bar{D}_s^{(*)}\rangle+\sqrt{\frac{2}{3}}|\Xi_c\bar{D}^{(*)}\rangle$\\
\hline
$|\bm{8},1,\frac{1}{2}\rangle$&$|\Lambda_c\bar{D}^{(*)}\rangle$\\
$|\bm{8},0,0\rangle$&$\sqrt{\frac{2}{3}}|\Lambda_c\bar{D}_s^{(*)}\rangle-\sqrt{\frac{1}{3}}|\Xi_c\bar{D}^{(*)}\rangle$\\
$|\bm{8},0,1\rangle$&$|\Xi_c\bar{D}^{(*)}\rangle$\\
$|\bm{8},-1,\frac{1}{2}\rangle$&$|\Xi_c\bar{D}_s^{(*)}\rangle$\\
\hline
$|\bm{8}^\prime,1,\frac{1}{2}\rangle$&$-|\Sigma_c^{(*)}\bar{D}^{(*)}\rangle$\\
$|\bm{8}^\prime,0,0\rangle$&$|\Xi_c^{\prime(*)}\bar{D}^{(*)}\rangle$\\
$|\bm{8}^\prime,0,1\rangle$&$\sqrt{\frac{2}{3}}|\Sigma^{(*)}_c\bar{D}^{(*)}_s\rangle-\sqrt{\frac{1}{3}}|\Xi_c^{\prime(*)}\bar{D}^{(*)}\rangle$\\
$|\bm{8}^\prime,-1,\frac{1}{2}\rangle$&$\sqrt{\frac{1}{3}}|\Xi_c^{\prime(*)}\bar{D}_{s}^{(*)}\rangle-\sqrt{\frac{2}{3}}|\Omega_c^{(*)}\bar{D}^{(*)}\rangle$\\
\hline
$|\bm{10},1,\frac{3}{2}\rangle$&$|\Sigma_c^{(*)}\bar{D}^{(*)}\rangle$\\
$|\bm{10},0,1\rangle$&$\sqrt{\frac{1}{3}}|\Sigma_c^{(*)}\bar{D}_s^{(*)}\rangle+\sqrt{\frac{2}{3}}|\Xi_c^{\prime(*)}\bar{D}^{(*)}\rangle$\\
$|\bm{10},-1,\frac{1}{2}\rangle$&$\sqrt{\frac{1}{3}}|\Omega_c^{(*)}\bar{D}^{(*)}\rangle+\sqrt{\frac{2}{3}}|\Xi_c^{\prime(*)}\bar{D}_s^{(*)}\rangle$\\
$|\bm{10},-2,0\rangle$&$|\Omega_c^{(*)}\bar{D}_s^{(*)}\rangle$\\
\bottomrule[1pt]
\end{tabular}\label{SU3ISO}
\end{table}

The spin wave functions of the considered single-charm baryons and anticharmed mesons are collected in Table \ref{spin WF}. For the $\Lambda_c$/$\Xi_c$ baryons, we first build a spin-$0$ light diquark and then couple it to the spin-$\tfrac12$ charm quark. Similarly, for the $\Sigma_c^{(*)}$/$\Xi_c^{\prime(*)}$/$\Omega_c^{(*)}$ baryons, we first construct a spin-$1$ light diquark and couple it to the spin-$\tfrac12$ charm quark to generate spin-$\tfrac12$ or spin-$\tfrac32$ spin wave functions.

\begin{table*}[htbp]
\setlength\tabcolsep{0.9pt} \caption{The spin wave functions of
the charmed mesons and baryons considered in this work. \label{spin WF}}
\begin{tabular}{|llc|llc|}
\toprule[1pt]
Hadron&$\left|Sm_S\right\rangle$&$\chi^{M}_{Sm_S}$&Hadron&$\left|Sm_S\right\rangle$&$\chi^{M}_{Sm_S}$\\
\hline
\multirow{3}{*}{$\bar{D}$/$\bar{D}_s$}&\multirow{3}{*}{$|00\rangle$}&\multirow{3}{*}{$\frac{1}{\sqrt{2}}(\uparrow\downarrow-\downarrow\uparrow)$}& \multirow{3}{*}{$\bar{D}^*$/$\bar{D}_s^*$}&$|11\rangle$&$\uparrow\uparrow$\\
&&&&$|10\rangle$&$\frac{1}{\sqrt{2}}(\uparrow\downarrow+\downarrow\uparrow)$\\
&&&&$|1-1\rangle$&$\downarrow\downarrow$\\
\hline
Hadron&$\left|Sm_S\right\rangle$&$\chi^{B}_{Sm_S}$&Hadron&$\left|Sm_S\right\rangle$&$\chi^{B}_{Sm_S}$\\
\hline
\multirow{2}{*}{$\Lambda_c$/$\Xi_c$}& $|\frac{1}{2}\frac{1}{2}\rangle$&$\frac{1}{\sqrt{2}}(\uparrow\downarrow-\downarrow\uparrow)\uparrow$&\multirow{4}{*}{$\Sigma_c^*$/$\Xi_c^*$/$\Omega_c^*$}&$|\frac{3}{2}\frac{3}{2}\rangle$&$\uparrow\uparrow\uparrow$\\
&$|\frac{1}{2}-\frac{1}{2}\rangle$&$\frac{1}{\sqrt{2}}(\uparrow\downarrow-\downarrow\uparrow)\downarrow$&&$|\frac{3}{2}\frac{1}{2}\rangle$&$\sqrt{\frac{1}{3}}(\uparrow\uparrow\downarrow+\uparrow\downarrow\uparrow+\downarrow\uparrow\uparrow)$\\
\multirow{2}{*}{$\Sigma_c$/$\Xi^\prime_c$/$\Omega_c$}&$|\frac{1}{2}\frac{1}{2}\rangle$&$-
\frac{1}{\sqrt{6}}(\uparrow\downarrow+\downarrow\uparrow)\uparrow+\sqrt{\frac{2}{3}}\uparrow\uparrow\downarrow$&&$|\frac{3}{2}-\frac{1}{2}\rangle$&$\sqrt{\frac{1}{3}}(\uparrow\downarrow\downarrow+\downarrow\uparrow\downarrow+\downarrow\downarrow\uparrow)$\\
&$|\frac{1}{2}-\frac{1}{2}\rangle$&$\frac{1}{\sqrt{6}}(\uparrow\downarrow+\downarrow\uparrow)\downarrow-\sqrt{\frac{2}{3}}\downarrow\downarrow\uparrow$&&$|\frac{3}{2}-\frac{3}{2}\rangle$&$\downarrow\downarrow\downarrow$\\
\bottomrule[1pt]
\end{tabular}
\end{table*}

The total wave function of the considered baryon-meson system with definite isospin $I$ and spin $J$ in the SU(3) basis can be written as
\begin{eqnarray}
\left|\left[BM\right]_{JJ_z}^{II_z}\right\rangle_{\text{SU(3)}}&=&|R,Y,I,I_z\rangle\nonumber\\
&&\otimes\sum_{m_{S_1},m_{S_2}}C_{S_1,m_{S_1};S_2,m_{S_2}}^{J,J_z}\chi^B_{S_1,m_{S_1}}\chi_{S_2,m_{S_2}}^M,\nonumber\\
\end{eqnarray}
where we explicitly present the third components of $I$ and $J$ with $I_z$ and $J_z$, respectively. Similarly, in the isospin basis, the total wave function can be written as
\begin{eqnarray}
\left|\left[BM\right]_{JJ_z}^{II_z}\right\rangle_{\text{iso}}&=&\sum_{m_{I_1}m_{I_2}}C_{I_1,m_{I_1};I_2,m_{I_2}}^{I,I_z}\phi_{I_1,m_{I_1}}^B\phi_{I_2,m_{I_2}}^M\nonumber\\
&&\otimes\sum_{m_{S_1},m_{S_2}}C_{S_1,m_{S_1};S_2,m_{S_2}}^{J,J_z}\chi^B_{S_1,m_{S_1}}\chi_{S_2,m_{S_2}}^M.\nonumber\\
\end{eqnarray}
The $C_{I_1,m_{I_1};I_2,m_{I_2}}^{I,I_z}$ and $C_{S_1,m_{S_1};S_2,m_{S_2}}^{J,J_z}$ are the SU(2) Clebsch-Gordan coefficients. The $\phi_{I_1,m_{I_1}}^B$ ($\chi^B_{S_1,m_{S_2}}$) and $\phi_{I_2,m_{I_2}}^M$ ($\chi^M_{S_2,m_{S_2}}$) are the flavor (spin) wave functions of the baryon and meson, respectively.

\section{Review the description of the observed $P_c$ and $P_{cs}$ states}\label{sec3}
In Sec. \ref{sec4}, we present a classification scheme for hidden-charm pentaquark states consisting of ($\Lambda_c$, $\Sigma_c^{(*)}$, $\Xi_c$, $\Xi_c^{\prime(*)}$, $\Omega_c^{(*)}$) baryons and ($\bar{D}^{(*)}$, $\bar{D}_s^{(*)}$) mesons. This scheme is inspired by our successful reproduction of the experimentally observed $P_c$ and $P_{cs}$ states, so we first recap our earlier characterization of these pentaquark states from previous works \cite{Chen:2022wkh,Chen:2024tuu}.

In Ref. \cite{Chen:2024tuu}, we assume the $P_c(4440)$ and $P_c(4457)$ as the $I=\frac{1}{2}$ $\Sigma_c\bar{D}^*$ molecular states with $J^P=\frac{1}{2}^-$ and $\frac{3}{2}^-$ for scenario 1 and $J^P=\frac{3}{2}^-$ and $\frac{1}{2}^-$ for scenario 2, respectively. For both scenarios, we adopt a dipole form factor with cutoff fixed at $1.0$ GeV and solve the coupled-channel Lippmann-Schwinger equation (LSE) by considering the $S$-wave channels consisting of the ($\Lambda_c$/$\Sigma_c^{(*)}$) and ($\bar{D}^{(*)}$) components. We use the experimental masses of $P_c(4440)$ and $P_{c}(4457)$ as inputs to extract the numerical values of $\tilde{g}_s$ and $\tilde{g}_a$, we obtain
\begin{eqnarray}
\text{Scenario}\, 1:&&\tilde{g}_s=8.28\,\text{GeV}^{-2},\,\tilde{g}_a=-1.46\,\text{GeV}^{-2},\label{scenario1}\\
\text{Scenario}\, 2:&&\tilde{g}_s=9.12\,\text{GeV}^{-2},\,\tilde{g}_a=1.25\,\text{GeV}^{-2}\label{scenario2}.
\end{eqnarray}

From Eqs. (\ref{scenario1}-\ref{scenario2}), we find that in both scenarios, the magnitude of $\tilde{g}_s$ associated with the operator $\bm{\lambda}_1\cdot\bm{\lambda}_2$ is much larger than that of $\tilde{g}_a$ associated with the operator $\bm{\lambda}_1\cdot\bm{\lambda}_2\,\bm{\sigma}_1\cdot\bm{\sigma}_2$. Accordingly, the sizable $\tilde{g}_s$ value points to a flavor-dominated selection rule governing interactions of all baryon-meson systems considered here. The $\tilde{g}_s\bm{\lambda}_1\cdot\bm{\lambda}_2$ term serves as the dominant attractive force generating possible bound states, whereas the $\tilde{g}_a\bm{\lambda}_1\cdot\bm{\lambda}_2\,\bm{\sigma}_1\cdot\bm{\sigma}_2$ term accounts for hyperfine splittings among baryon-meson systems.

Accordingly, we may straightforwardly evaluate matrix elements
\begin{eqnarray}
\left\langle\bm{\lambda}_1\cdot\bm{\lambda}_2\right\rangle_{\text{SU(3)}}=\left\langle R,Y,I\left|\bm{\lambda}_1\cdot\bm{\lambda}_2\right|R,Y,I\right\rangle\label{SU3FLL}
\end{eqnarray}
to extract basic interaction properties arising from the flavor singlet $\mathbf{1}$, octet ($\mathbf{8}$/$\mathbf{8}^\prime$), and decuplet $\mathbf{10}$ representations.
\begin{table}[htbp]
\renewcommand\arraystretch{1.5}
\caption{The matrix elements $\langle\bm{\lambda}_1\cdot\bm{\lambda}_2\rangle_{\text{SU(3)}}$ (defined in Eq. (\ref{SU3FLL})) of the flavor $\bm{1}$, $\bm{8}$, $\bm{8}^\prime$, and $\bm{10}$ representations.}
\begin{tabular}{c|cccccccccccccc}
\toprule[0.8pt]
&$|\bm{1}\rangle$&$|\bm{8}\rangle$&$|\bm{8}^\prime\rangle$&$|\bm{10}\rangle$\\
\hline
$\langle\bm{\lambda}_1\cdot\bm{\lambda}_2\rangle_{\text{SU(3)}}$&$-\frac{16}{3}$&$\frac{2}{3}$&$-\frac{10}{3}$&$\frac{8}{3}$\\
\hline
\end{tabular}\label{SU3LL}
\end{table}

The results of $\langle\bm{\lambda}_1\cdot\bm{\lambda}_2\rangle_{\text{SU(3)}}$ for the $|\bm{1}\rangle$, $|\bm{8}\rangle$, $|\bm{8}^\prime\rangle$, and $|\bm{10}\rangle$ are listed in Table \ref{SU3LL}. Since the $\tilde{g}_s$ has positive values in both scenarios, from Table \ref{SU3LL}, we conclude that the states that belong to the $\bm{1}$ and $\bm{8}^\prime$ representations have attractive forces due to their negative $\langle\bm{\lambda}_1\cdot\bm{\lambda}_2\rangle_{\text{SU(3)}}$ matrix elements, whereas those transforming as $\bm{8}$ and $\bm{10}$ have repulsive forces due to their positive $\langle\bm{\lambda}_1\cdot\bm{\lambda}_2\rangle_{\text{SU(3)}}$ matrix elements.

As listed in Table \ref{SU3ISO}, the triple-strange hidden-charm system $\Omega_c^{(*)}\bar{D}_s^{(*)}$ has the SU(3) flavor wave function $|\bm{10},-2,0\rangle$, the three light $s$ quark are fully symmetric, therefore, this selection rule prohibits the existences of any triple-strange hidden-charm pentaquarks.

Here we remind the readers that our conclusion regarding the absence of triple-strange hidden-charm pentaquarks is model-dependent. For instance, the $\Omega_c^{(*)}\bar{D}_s^{(*)}$ system was also explored within the one-boson-exchange model \cite{Wang:2021hql}. After incorporating the $S-D$ wave mixing and coupled-channel effects, the authors suggested that the $\Omega_c\bar{D}_s^*$ state with $J^P=\frac{3}{2}^-$ and the $\Omega_c^*\bar{D}_s^*$ state with $J^P=\frac{5}{2}^-$ may form bound states with cutoff values slightly above 1.6 GeV.

In the light-meson exchange saturation model \cite{Yang:2022ezl}, the authors introduced the contact potentials generated by the scalar and vector light mesons to describe the meson-baryon interactions. They suggested that there may exist several decuplet hidden-charm pentaquarks including the triple-strange hidden-charm molecules. The $P_{c\bar{c}sss}$ states have also been investigated in Ref. \cite{Roca:2024nsi} within the local hidden gauge symmetry approach, they found that the considered meson-baryon interaction is not enough to generate a bound or resonance state. However, in Ref. \cite{Clymton:2025dzt}, the authors suggested that if they further consider the $\Omega_c\bar{D}_s$, $\Omega_c\bar{D}_s^*$, and $\Omega_c^*\bar{D}_s^*$ channels, they can find two triple-strange hidden-charm pentaquarks with $J^P=\frac{1}{2}^-$, i.e., the $P_{c\bar{c}sss}(4787)$ and $P_{c\bar{c}sss}(4841)$ states. Additionally, Ref. \cite{Meng:2019fan} investigated $P_{c\bar{c}sss}$ states within the constituent quark model and identified four compact pentaquark resonance candidates carrying triple strange quantum numbers.

Returning to our model, we restrict our discussion to channels with attractive interactions, namely those transforming as the $\mathbf{1}$ and $\mathbf{8}^\prime$ representations. Explicitly, Table \ref{SU3ISO} lists the relevant eigenstates: the $|\mathbf{1},0,0\rangle$ singlet, $|\mathbf{8}^\prime,1,\tfrac12\rangle$ doublet, $|\mathbf{8}^\prime,0,0\rangle$ singlet, $|\mathbf{8}^\prime,0,1\rangle$ triplet, and $|\mathbf{8}^\prime,-1,\tfrac12\rangle$ doublet.

As collected in Table \ref{SU3ISO}, the SU(3)-basis states $|\mathbf{8}^\prime,1,\tfrac12\rangle$ and $|\mathbf{8}^\prime,0,0\rangle$ exactly match the isospin-basis wave functions of the $I=\tfrac12$ $\Sigma_c^{(*)}\bar{D}^{(*)}$ system and $I=0$ $\Xi_c^{\prime(*)}\bar{D}^{(*)}$ system, respectively. In contrast, the SU(3) eigenstates $|\mathbf{1},0,0\rangle$, $|\mathbf{8}^\prime,0,1\rangle$, and $|\mathbf{8}^\prime,-1,\tfrac12\rangle$ take the forms of linear superpositions of three sets of isospin-basis configurations: $(\Lambda_c\bar{D}_s^{(*)},\,\Xi_c\bar{D}^{(*)})$, $(\Sigma_c^{(*)}\bar{D}_s^{(*)},\,\Xi_c^{\prime(*)}\bar{D}^{(*)})$, and $(\Xi_c^{\prime(*)}\bar{D}_s^{(*)},\,\Omega_c^{(*)}\bar{D}^{(*)})$, respectively.

Thus, on the one hand, for the $|\mathbf{8}^\prime,1,\tfrac12\rangle$ and $|\mathbf{8}^\prime,0,0\rangle$ states, their expressions of flavor wave functions are identical in the SU(3) and isospin bases, so we can calculate their molecular spectra using flavor wave functions from either basis. In Ref. \cite{Chen:2024tuu}, we use the parameters in both scenarios to predict the possible molecular spectra of the $|\bm{8}^\prime,1,\frac{1}{2}\rangle$ ($\Sigma_c^{(*)}\bar{D}^{(*)}$ with $I=\frac{1}{2}$) and $|\bm{8}^\prime,0,0\rangle$ ($\Xi_c^{\prime(*)}\bar{D}^{(*)}$ with $I=0$) states. Since their interactions are both from the exchanges of isospin singlet light mesons and isospin triplet light mesons, so their spectra are very similar to each other. 

On the other hand, the $|\mathbf{1},0,0\rangle$, $|\mathbf{8}^\prime,0,1\rangle$, and $|\mathbf{8}^\prime,-1,\tfrac12\rangle$ eigenstates cannot be directly mapped onto physical eigenstates. Owing to SU(3) flavor breaking induced by the mass difference between strange $s$ quarks and light $u/d$ quarks, the presence of an $s$ quark inside either the single-charm baryon or anticharmed meson splits the threshold into two distinct baryon-meson mass thresholds.

To analyze the mass spectrum of the broken $|\mathbf{1},0,0\rangle$ singlet, which mixes the physical $\Lambda_c\bar{D}_s^{(*)}$ and $\Xi_c\bar{D}^{(*)}$ components, we examined the mass of $P_{cs}(4338)$ through channel mixing between $\Lambda_c\bar{D}_s$ and $\Xi_c\bar{D}$ in Refs. \cite{Chen:2022wkh,Chen:2024tuu}. We briefly recap our earlier results here, which would be helpful to clarify our classification scheme for the considered hidden-charm pentaquarks.

The effective potential matrix of the $J=\frac{1}{2}$ ($\Lambda_c\bar{D}_s$, $\Xi_c\bar{D}$) can be written as
\begin{eqnarray}
\mathbb{V}&=&\left(
             \begin{array}{cc}
               V^{\Lambda_c\bar{D}_s\rightarrow\Lambda_c\bar{D}_s} & V^{\Lambda_c\bar{D}_s\rightarrow \Xi_c\bar{D}} \\
               V^{\Xi_c\bar{D}\rightarrow\Lambda_c\bar{D}_s} & V^{\Xi_c\bar{D}\rightarrow\Xi_c\bar{D}} \\
             \end{array}
           \right)\nonumber\\
           &=&\left(
             \begin{array}{cc}
              -\frac{4}{3}\tilde{g}_s  & -2\sqrt{2}\tilde{g}_sg_x \\
              -2\sqrt{2}\tilde{g}_sg_x  & -\frac{10}{3}\tilde{g}_s \\
             \end{array}
           \right).\label{VLX}
\end{eqnarray}

As discussed in Sec. \ref{sec21}, the $\Lambda_c\bar{D}_s$ and $\Xi_c\bar{D}$ channels interact through the exchange of strange light mesons originating from the $\sum_{j=4}^7\lambda_1^j\lambda_2^j$ operator. We therefore introduce a scaling factor $g_x$ for the off-diagonal matrix element $V^{\Lambda_c\bar{D}_s\to \Xi_c\bar{D}}$ (and its counterpart $V^{\Xi_c\bar{D}\to\Lambda_c\bar{D}_s}$) to parameterize the suppression effects induced by strange-meson exchange.

\begin{figure*}[!htbp]
    \centering
    \includegraphics[width=0.8\linewidth]{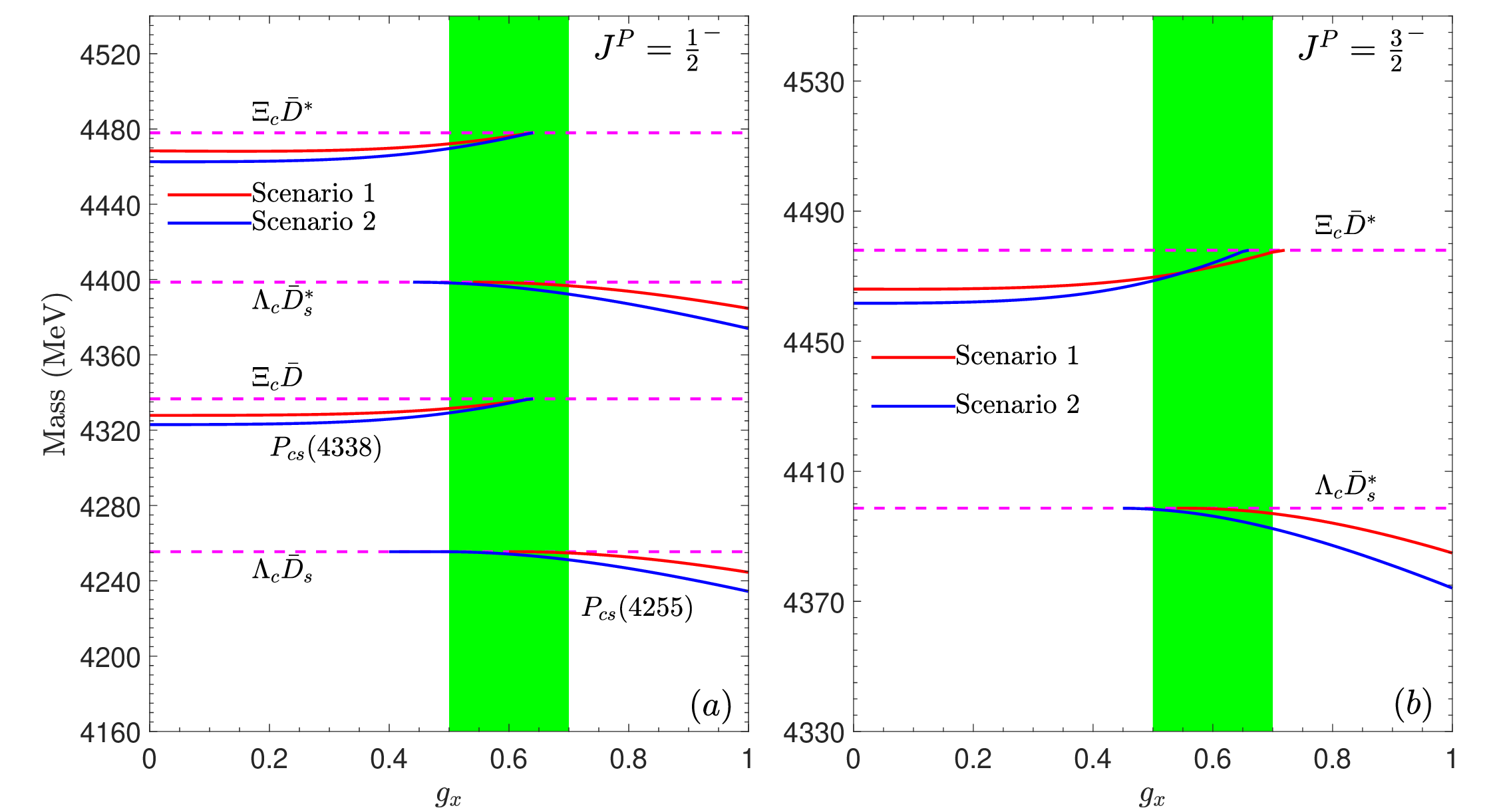}
    \caption{The $g_x$ dependences of the pole positions produced from the $\Lambda_c\bar{D}_s^{(*)}-\Xi_c\bar{D}^{(*)}$ mixing with $J^P=\frac{1}{2}^-$ (a) and $\frac{3}{2}^-$ (b). The pole trajectories obtained from scenario 1 and scenario 2 are illustrated with red and blue lines, respectively. The physical region with $g_x=0.6\pm0.1$ are illustrated with green bands.}
    \label{Pcs}
\end{figure*}

As given in Eq. (\ref{VLX}), the $\Xi_c\bar{D}\rightarrow \Xi_c\bar{D}$ channel has a relatively strong attractive force with $-\frac{10}{3}\tilde{g}_s$, whereas the $\Lambda_c\bar{D}_s\rightarrow\Lambda_c\bar{D}_s$ channel has a weak attractive force with $-\frac{4}{3}\tilde{g}_s$. The factor $g_x$ is restricted to the interval [0,1]. In Figs. \ref{Pcs} (a) and (b), by varying the $g_x$ from 0 to 1, we obtain the $g_x$-dependences of the pole positions that are close to the $\Lambda_c\bar{D}_s$ and $\Xi_c\bar{D}$ thresholds with $J^P=\frac{1}{2}^-$ and $\frac{3}{2}^-$, respectively. The results calculated with the parameters from scenario 1 and scenario 2 are labelled with red and blue lines, respectively. 

As displayed in Fig. \ref{Pcs}, when $g_x$ rises from $0$ to approximately $0.6$, the pole corresponding to $P_{cs}(4338)$ shifts toward the $\Xi_c\bar{D}$ threshold, while as $g_x$ increases from roughly 0.6 to 1, the pole of $P_{cs}(4255)$ moves away from the threshold of $\Lambda_c\bar{D}_s$. This behavior indicates that the channel mixing induced from the SU(3) breaking effect redistributes the attractive forces among the $\Xi_c\bar{D}\rightarrow \Xi_c\bar{D}$ and $\Lambda_c\bar{D}_s\rightarrow\Lambda_c\bar{D}_s$ channels.

We also point out that the pole behaviors arising from the $\Lambda_c\bar{D}_s$-$\Xi_c\bar{D}$ coupling are also model-dependent. For instance, Ref. \cite{Feijoo:2022rxf} also studied this mixing effect. The authors found that incorporating the $\Lambda_c\bar{D}_s$ channel generates additional attraction in the $\Xi_c\bar{D}$ channel, a prediction opposite to ours. Accordingly, further experimental data are required to pin down the actual impact of $\Lambda_c\bar{D}_s$-$\Xi_c\bar{D}$ channel mixing.

The $P_{cs}(4459)$ is reported by the LHCb Collaboration \cite{LHCb:2020jpq} with $m_{P_{cs}(4459)}=4458.8^{+6.0}_{-3.1}$ MeV, then the Belle Collaboration \cite{Belle:2025pey} reported the evidence supporting the existence of $P_{cs}(4459)$ but the measured mass is $4471.7\pm4.8\pm0.6$ MeV. Although Ref. \cite{Clymton:2025hez} suggest that the $P_{cs}(4459)$ and $P_{cs}(4472)$ might be different states with $J^P=\frac{3}{2}^-$ and $\frac{1}{2}^-$. From our model, the obtained mass of the $\Xi_c\bar{D}$ state with $J=\frac{1}{2}$ and $\frac{3}{2}$ are degenerate, since the matrix element of $\langle\tilde{g}_a\bm{\lambda}_1\cdot\bm{\lambda}_2\bm{\sigma}_1\cdot\bm{\sigma}_2\rangle$ for the $J=\frac{1}{2}$ and $\frac{3}{2}$ $\Xi_c\bar{D}^*$ states are both 0. Note that these two states are both observed in the $J/\Psi\Lambda^0$ final state with $I=0$, and the $\Xi_c\bar{D}^*$ (4478 MeV) threshold is the nearest threshold to these two states, thus, we tentatively assume that the observed $P_{cs}(4459)$ or $P_{cs}(4472)$ state is a molecular state consists of $\Xi_c\bar{D}^*$ component. As shown in Figs. \ref{Pcs} (a) and (b), in both scenarios, the $\Lambda_c\bar{D}_s^*-\Xi_c\bar{D}^*$ coupling also plays similar role to that of the $\Lambda_c\bar{D}_s-\Xi_c\bar{D}$ coupling. Thus, the flavor of the $P_{cs}(4338)$ and $P_{cs}(4459)$/$P_{cs}(4472)$ all belong to the broken $|\bm{1},0,0\rangle$ singlet.

To estimate the existences of the possible bound states that belong to the broken $|\bm{8}^\prime,0,1\rangle$ triplet and broken $|\bm{8}^\prime,-1,\frac{1}{2}\rangle$ doublet, we need to constrain the $g_x$ value. Note that the reported mass of the $P_{cs}(4338)$ \cite{LHCb:2022ogu} is very close to the threshold of the $\Xi_c\bar{D}$, thus, from the pole trajectories illustrated in both scenarios in Fig. \ref{Pcs} (a), we set $g_x=0.6\pm0.1$ as the physical region to reproduce the mass of the $P_{cs}(4338)$ state, and we highlight this region with green bands in Figs. \ref{Pcs} (a) and (b).

\section{Classifying the hidden charm pentaquark states}\label{sec4}
\begin{figure*}[!htbp]
    \centering
    \includegraphics[width=0.7\linewidth]{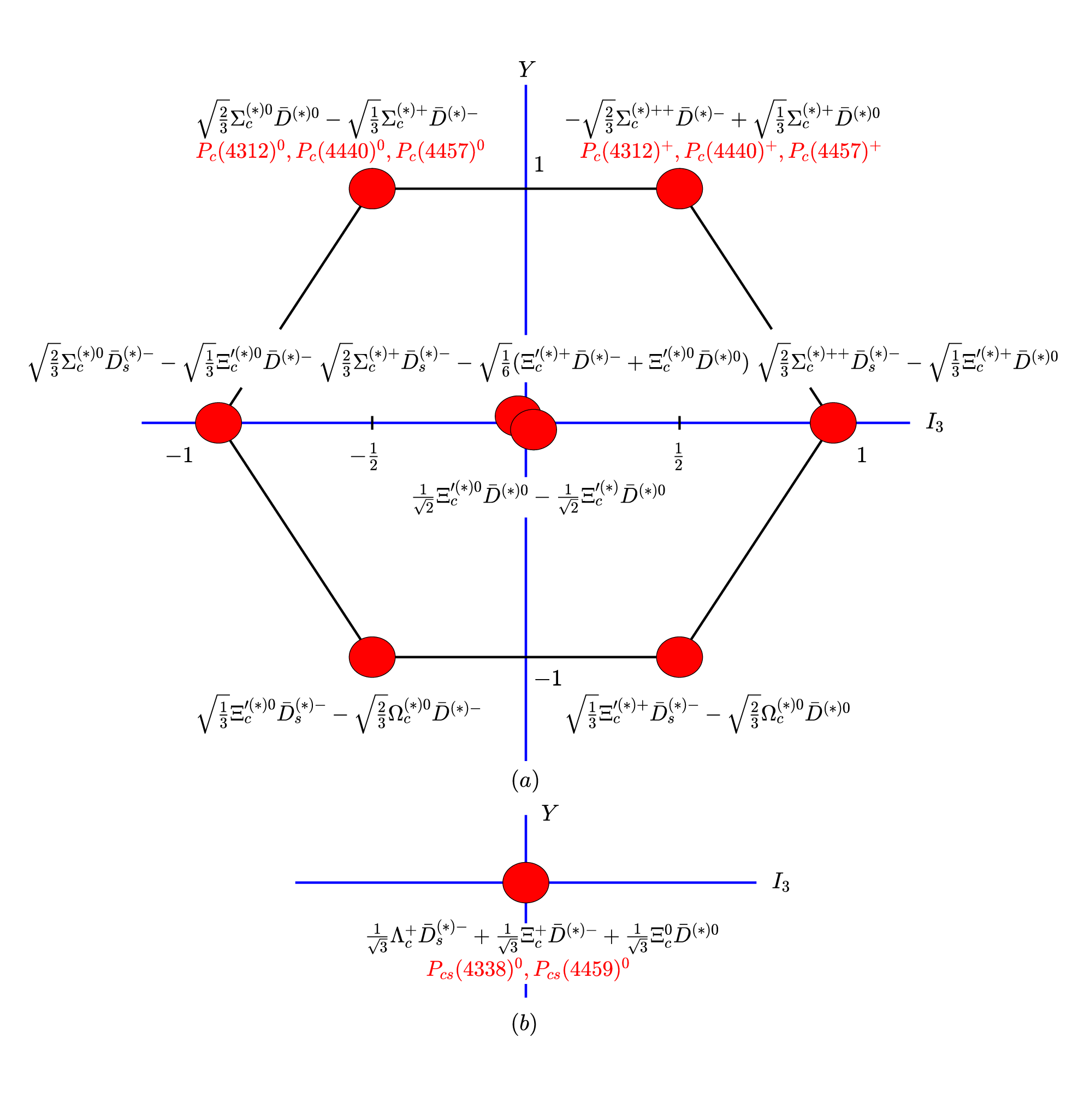}
    \caption{The flavor wave functions of molecular hidden-charm pentaquarks constructed in the SU(3) basis that belong to the $\bm{8}^\prime$ (a) and $\bm{1}$ (b) representations.}
    \label{Oct}
\end{figure*}

The above satisfactory description of the observed $P_c$ and $P_{cs}$ states inspires us to propose the picture illustrated in Fig. \ref{Oct} to classify the hidden-charm pentaquark states consisting of the single-charm baryons and anticharmed mesons. In Figs. \ref{Oct} (a) and (b), we plot the flavor wave functions of the hidden charm pentaquark states belong to the $\bm{8}^\prime$ and $\bm{1}$ representations, respectively. Due to their negative matrix elements $\langle\bm{\lambda}_1\cdot\bm{\lambda}_2\rangle_{\text{SU(3)}}$ listed in Table \ref{SU3LL} and the dominant positive coupling $\tilde{g}_s$ extracted from the masses of $P_c(4440)$ and $P_c(4457)$ in both scenarios in Eqs. (\ref{scenario1}-\ref{scenario2}), we can expect that only the state that belong to the $\bm{8}^\prime$ and $\bm{1}$ representations can form bound states. Oppositely, the states that belong to the $\bm{8}$ and $\bm{10}$ representations can not form bound states.

By checking the available experimental information, our classification method in Figs. \ref{Oct} (a) and (b) can be summarized as follows
\begin{itemize}
\item[(1)]The SU(3) flavor wave function $|\bm{R},Y,I,I_z\rangle\equiv|\bm{8}^\prime,1,\frac{1}{2},\frac{1}{2}\rangle$ can correspond to the observed $P_c(4312)^+$, $P_{c}(4440)^+$, and $P_{c}(4457)^+$ states, these three states are observed in the $J/\Psi p^+$ final state \cite{LHCb:2015yax,LHCb:2016ztz,LHCb:2019kea}.
\item[(2)]The flavor wave function $|\bm{8}^\prime,1,\frac{1}{2},-\frac{1}{2}\rangle$ corresponds to the neutral $P_c(4312)^0$, $P_c(4440)^0$, and $P_c(4457)^0$ states, these states have not been reported yet.
\item[(3)]The flavor wave function $|\bm{8}^\prime,0,0\rangle$ coincides with that of the $I=0$ $\Xi_c^{\prime(*)}\bar{D}^{(*)}$ system. Such pentaquark candidate has not been reported yet.
\item[(4)]The SU(3) flavor wave functions $|\bm{8}^\prime,0,1\rangle$ with $I_z=1$, $0$, and $-1$ do not correspond to pure physical states, they reduce to the mixture of the physical $\Sigma_c^{(*)}\bar{D}_s^{(*)}$ and $\Xi_c^{\prime(*)}\bar{D}^{(*)}$ components. The $\Sigma_c^{(*)}\bar{D}_s^{(*)}$ or $\Xi_c^{\prime(*)}\bar{D}^{(*)}$ molecular states have not been observed yet.
\item[(5)]The SU(3) flavor wave functions $|\bm{8}^\prime,-1,\frac{1}{2}\rangle$ with $I_3=\pm\frac{1}{2}$ do not correspond to pure physical states, they reduce to the mixture of the physical $\Xi_c^{\prime(*)}\bar{D}_s^{(*)}$ and $\Omega_c^{(*)}\bar{D}^{(*)}$ components. The $\Xi_c^{\prime(*)}\bar{D}_s^{(*)}$ and $\Omega_c^{(*)}\bar{D}^{(*)}$ molecular states have not been observed yet.
\item[(6)]The SU(3) flavor wave function $|\bm{1},0,0\rangle$ does not correspond to pure physical state, it reduce to the mixture of the physical $\Lambda_c\bar{D}_s^{(*)}$ and $\Xi_c\bar{D}^{(*)}$ components, the $\Xi_c\bar{D}$ candidate $P_{cs}(4338)$ \cite{LHCb:2022ogu} and the $\Xi_c\bar{D}^*$ candidate $P_{cs}(4459)$/$P_{cs}(4472)$ \cite{LHCb:2020jpq,Belle:2025pey} are reported in the $J/\Psi\Lambda^0$ final states.
\end{itemize}

Among the items (1)-(6), the items (1), (2), (3), and (6) have been investigated in Ref. \cite{Chen:2024tuu} with the parameters $\tilde{g}_s$ and $\tilde{g}_a$ extracted from both scenarios, one can refer to Ref. \cite{Chen:2024tuu} for more detail.

Our classification method naturally leads to the SU(3) $|\bm{8}^\prime, 0,1\rangle$ and $|\bm{8}^\prime,-1,\frac{1}{2}\rangle$ molecular states suggested in items (4) and (5). However, SU(3) flavor breaking splits each pure SU(3) eigenstate into a mixture of two coupled channels, i.e., $\Sigma_c^{(*)}\bar{D}_s^{(*)}-\Xi_c^{\prime(*)}\bar{D}^{(*)}$ for $|\mathbf{8}^\prime,0,1\rangle$ and $\Xi_c^{\prime(*)}\bar{D}_s^{(*)}-\Omega_c^{(*)}\bar{D}^{(*)}$ for $|\mathbf{8}^\prime,-1,\tfrac12\rangle$. The mixing patterns in items (4)-(5) are very similar to that of item (6). In item (6), we consider the $\Lambda_c\bar{D}_s-\Xi_c\bar{D}$ mixing to explain why the mass of $P_{cs}(4338)$ is so close to the $\Xi_c\bar{D}$ threshold. Thus, following the same coupled-channel formalism used to analyze $P_{cs}(4338)$, we carry out analogous coupled-channel calculations to explore the mixing effects within the broken $|\mathbf{8}^\prime,0,1\rangle$ and $|\mathbf{8}^\prime,-1,\tfrac12\rangle$ SU(3) states from items (4) and (5). Furthermore, the mixing between $\Xi_c^{\prime(*)}\bar{D}_s^{(*)}$ and $\Omega_c^{(*)}\bar{D}^{(*)}$ has been emphasized in Refs. \cite{Marse-Valera:2022khy,Roca:2024nsi}. Those authors demonstrated that off-diagonal interaction terms are critical for generating sufficient attractive forces.

\begin{figure*}[!htbp]
    \centering
    \includegraphics[width=0.8\linewidth]{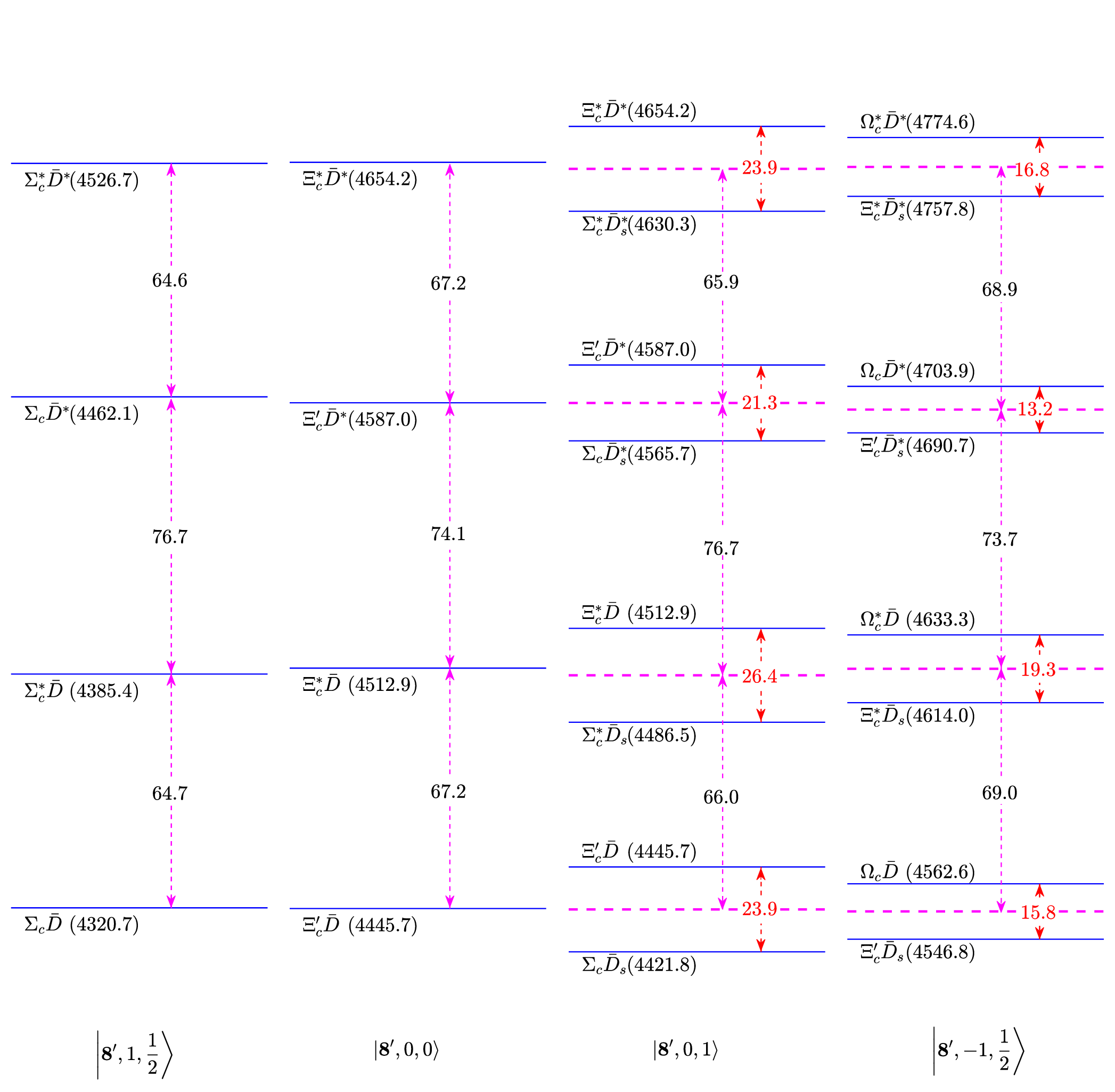}
    \caption{The baryon-meson thresholds related to the $|\bm{8}^\prime,1,\frac{1}{2}\rangle$, $|\bm{8}^\prime,0,0\rangle$, $|\bm{8}^\prime,0,1\rangle$, and $|\bm{8}^\prime,-1,\frac{1}{2}\rangle$ flavor wave functions.}
    \label{Octth}
\end{figure*}

In Fig. \ref{Octth}, we collect the thresholds that are relevant to the $|\bm{8}^\prime,1,\frac{1}{2}\rangle$, $|\bm{8}^\prime,0,0\rangle$, $|\bm{8}^\prime,0,1\rangle$, and $|\bm{8}^\prime,-1,\frac{1}{2}\rangle$ states. As shown in Fig. \ref{Octth}, the mass gaps of the
\begin{eqnarray}
\Sigma_c^*\bar{D}^*-\Sigma_c\bar{D}^*-\Sigma_c^*\bar{D}-\Sigma_c\bar{D}
\end{eqnarray}in $|\bm{8}^\prime,1,\frac{1}{2}\rangle$ are very similar to that of the
\begin{eqnarray}
\Xi_c^*\bar{D}^*-\Xi_c^\prime\bar{D}^*-\Xi_c^*\bar{D}-\Xi_c^\prime\bar{D}
\end{eqnarray}
in $|\bm{8}^\prime,0,0\rangle$. For the thresholds that are relevant to the $|\bm{8}^\prime,0,1\rangle$ and $|\bm{8}^\prime,-1,\frac{1}{2}\rangle$, such similarity reduce to the mass gaps of the
\begin{eqnarray}
&&(\Xi_c^*\bar{D}^*,\Sigma_c^*\bar{D}_s^*)_A-(\Xi_c^\prime\bar{D}^*,\Sigma_c\bar{D}_s^*)_A\nonumber\\&-&(\Xi_c^*\bar{D},\Sigma_c^*\bar{D}_s)_A-(\Xi_c^\prime\bar{D},\Sigma_c\bar{D}_s)_A \end{eqnarray}
and
\begin{eqnarray}
&&(\Omega^*_c\bar{D}^*,\Xi_c^*\bar{D}_s^*)_A-(\Omega_c\bar{D}^*,\Xi_c^\prime\bar{D}_s^*)_A\nonumber\\&-&(\Omega_c^*\bar{D},\Xi_c^*\bar{D}_s)_A-(\Omega_c\bar{D},\Xi_c^\prime\bar{D}_s)_A,
\end{eqnarray}
respectively, where we use the notation $(\Xi_c^*\bar{D}^*,\Sigma_c^*\bar{D}_s^*)_A$ to denote the average threshold of the $\Xi_c^*\bar{D}^*$ and $\Sigma_c^*\bar{D}_s^*$ channels, the average values are illustrated with blue-dotted lines in Fig. \ref{Octth}.


\section{Investigating the possible $\Sigma_c^{(*)}\bar{D}^{(*)}_s-\Xi_c^{\prime(*)}\bar{D}^{(*)}$ and $\Xi_c^{\prime(*)}\bar{D}_s^{(*)}-\Omega_c^{(*)}\bar{D}^{(*)}$ bound states}\label{sec5}
In Ref. \cite{Chen:2022wkh}, we performed the single-channel and coupled-channel calculations to the $I=\frac{1}{2}$ $\Sigma_c^{(*)}\bar{D}^{(*)}$  and $I=0$ $\Xi_c^{\prime(*)}\bar{D}^{(*)}$ systems, whose flavor wave functions exactly match the SU(3) eigenstates $|\bm{8}^\prime,1,\frac{1}{2}\rangle$ and $|\bm{8}^\prime,0,0\rangle$, respectively. Our calculation shows that for the bound state below the threshold of $\Sigma_c\bar{D}$, $\Sigma_c^*\bar{D}$, $\Sigma_c\bar{D}^*$, or $\Sigma_c^*\bar{D}^*$ with $J^P=\frac{1}{2}^-$ or $\frac{3}{2}^-$, the coupled-channel corrections originating from the remaining $\Sigma^{(*)}_c\bar{D}^{(*)}$ channels are very small, since these corrections stem solely from off-diagonal spin-spin matrix elements $\tilde{g}_a\langle\bm{\lambda}_1\cdot\bm{\lambda}_2\bm{\sigma}_1\cdot\bm{\sigma}_2\rangle$.

Similarly, the bound state below the threshold of $\Xi^\prime_c\bar{D}$, $\Xi_c^*\bar{D}$, $\Xi^\prime_c\bar{D}^*$, or $\Xi_c^*\bar{D}^*$ with $J^P=\frac{1}{2}^-$ or $\frac{3}{2}^-$ also receives minor corrections from the coupled-channel effects introduced from the rest of $\Xi^{\prime(*)}_c\bar{D}^{(*)}$ channels. By comparing the obtained masses of the $I=\frac{1}{2}$ $\Sigma_c^{(*)}\bar{D}^{(*)}$  and $I=0$ $\Xi_c^{\prime(*)}\bar{D}^{(*)}$ bound states, we conclude that the coupled-channel corrections to the masses of the considered bound states are within 0-5 MeV. Such modest shifts cannot alter the spin assignments, which are primarily governed by the diagonal spin-spin matrix elements $\tilde{g}_a\langle\bm{\lambda}_1\cdot\bm{\lambda}_2\bm{\sigma}_1\cdot\bm{\sigma}_2\rangle$.

In contrast to the coupled-channel corrections for the $|\mathbf{8}^\prime,1,\tfrac12\rangle$ and $|\mathbf{8}^\prime,0,0\rangle$ multiplets, which originate from off-diagonal spin-spin interaction terms $\tilde{g}_a\bm{\lambda}_1\cdot\bm{\lambda}_2\bm{\sigma}_1\cdot\bm{\sigma}_2$, the off-diagonal mixing corrections for $\Lambda_c\bar{D}_s^{(*)}-\Xi_c\bar{D}^{(*)}$ coupling arise from the flavor mixing terms $\tilde{g}_s\bm{\lambda}_1\cdot\bm{\lambda}_2$. As illustrated in Fig. \ref{Pcs}, such corrections significantly shift the masses of the $P_{cs}(4338)$ and $P_{cs}(4459)$.

Besides the SU(3)-broken $|\mathbf{1},0,0\rangle$ singlet, which decomposes into a linear admixture of physical $\Lambda_c\bar{D}_s^{(*)}$ and $\Xi_c\bar{D}^{(*)}$ channels, the $|\mathbf{8}^\prime,0,1\rangle$ and $|\mathbf{8}^\prime,-1,\tfrac12\rangle$ eigenstates also receive flavor-mixing contributions from off-diagonal matrix elements $\tilde{g}_s\langle\bm{\lambda}_1\cdot\bm{\lambda}_2\rangle$. These mixing terms correspond to the $\Sigma_c^{(*)}\bar{D}_s^{(*)}-\Xi_c^{\prime(*)}\bar{D}^{(*)}$ and $\Xi_c^{\prime(*)}\bar{D}_s^{(*)}-\Omega_c^{(*)}\bar{D}^{(*)}$ coupled channels, respectively.

Now we aim to explore potential bound states corresponding to the flavor eigenstates $|\mathbf{8}^\prime,0,1\rangle$ and $|\mathbf{8}^\prime,-1,\tfrac12\rangle$. Once SU(3) symmetry breaking is incorporated, these two multiplets decompose into the coupled-channel systems $\Sigma_c^{(*)}\bar{D}_s^{(*)}-\Xi_c^{\prime(*)}\bar{D}^{(*)}$ and $\Xi_c^{\prime(*)}\bar{D}_s^{(*)}-\Omega_c^{(*)}\bar{D}^{(*)}$, respectively.

To decide whether additional coupled channels need to be considered, we draw on our earlier results for the $|\mathbf{8}^\prime,1,\tfrac12\rangle$ and $|\mathbf{8}^\prime,0,0\rangle$ states. Those calculations show that mixing corrections originating from the off-diagonal spin-spin operator $\tilde{g}_a\bm{\lambda}_1\cdot\bm{\lambda}_2\bm{\sigma}_1\cdot\bm{\sigma}_2$ are negligible. We may therefore safely ignore such contributions and retain only channel mixing generated by the off-diagonal flavor mixing term $\tilde{g}_s\bm{\lambda}_1\cdot\bm{\lambda}_2$, namely the two-channel couplings $\Sigma_c^{(*)}\bar{D}_s^{(*)}-\Xi_c^{\prime(*)}\bar{D}^{(*)}$ and $\Xi_c^{\prime(*)}\bar{D}_s^{(*)}-\Omega_c^{(*)}\bar{D}^{(*)}$.

The mixtures of the $\Sigma_c^{(*)}\bar{D}_s^{(*)}$-$\Xi_c^{\prime(*)}\bar{D}^{(*)}$ and $\Xi_c^{\prime(*)}\bar{D}_s^{(*)}-\Omega_c^{(*)}\bar{D}^{(*)}$ have already been stressed in Ref. \cite{Peng:2019wys}, by considering the heavy-flavor symmetry and SU(3) flavor symmetry, they predicted several pentaquark candidates. The resonance saturation model was adopted in Ref. \cite{Yang:2022ezl}, they introduced the potentials that are saturated by the scalar and vector light mesons and found that the $\Sigma_c^{(*)}\bar{D}_s^{(*)}$-$\Xi_c^{\prime(*)}\bar{D}^{(*)}$ and $\Xi_c^{\prime(*)}\bar{D}_s^{(*)}$-$\Omega^{(*)}_c\bar{D}^{(*)}$ may lead to various bound/virtual/ resonance states. Within the local hidden gauge formalism \cite{Garcia-Gonzales:2026dnn}, the authors also investigated the mixture of $\Sigma_c\bar{D}_s$-$\Xi_c^{\prime}\bar{D}$, their results also supported the emergences of $P_{cs}$ states with $I=1$. A constituent quark model calculation was performed in Ref. \cite{Ortega:2022uyu}, by considering the $\Sigma_c^{(*)}\bar{D}_s^{(*)}$-$\Xi_c^{\prime(*)}\bar{D}^{(*)}$ mixing and some of other channels, they found the $I(J^P)=1(\frac{3}{2}^-)$ $P_{cs}(4547)$ and $I(J^P)=1(\frac{5}{2}^-)$ $P_{cs}(4456)$ pentaquark states.

In Table \ref{matrixele}, we list the matrix elements
\begin{eqnarray}
\langle\mathcal{O}^{\text{f}}\rangle&=&\langle [B_1M_1]_{JJ_z}^{II_z}|\bm{\lambda}_1\cdot\bm{\lambda}_2|[B_2M_2]_{JJ_z}^{II_z}\rangle,\\
\langle\mathcal{O}^{\text{fs}}\rangle&=&\langle[B_1M_1]_{JJ_z}^{II_z}|\bm{\lambda}_1\cdot\bm{\lambda}_2\bm{\sigma}_1\cdot\bm{\sigma}_2|[B_2M_2]_{JJ_z}^{II_z}\rangle,
\end{eqnarray}
of the single-strange hidden-charm $\Sigma_c^{(*)}\bar{D}_s^{(*)}-\Xi_c^{\prime(*)}\bar{D}^{(*)}$ systems and double-strange hidden-charm $\Xi_c^{\prime(*)}\bar{D}_s^{(*)}-\Omega_c^{(*)}\bar{D}^{(*)}$ systems with $J^P=\frac{1}{2}^-$, $\frac{3}{2}^-$, and $\frac{5}{2}^-$. Here, since the off-diagonal terms are from the exchanges of strange light mesons, i.e., the $\sum_{j=4}^7\lambda_1^j\lambda_2^j$, we multiply the factor $g_x$ to take into account the SU(3) breaking effect.

As shown in Table \ref{SU3ISO}, since the mixture of $\Sigma_c^{(*)}\bar{D}_s^{(*)}$ ($\Xi_c^{\prime(*)}\bar{D}_s^{(*)}$) and $\Xi_c^{\prime(*)}\bar{D}^{(*)}$ ($\Omega_c^{(*)}\bar{D}^{(*)}$) can reproduce the SU(3) eigenstates $|\bm{8}^\prime,0,1\rangle$ and $|\bm{10},0,1\rangle$, thus the eigenvalues listed in Table \ref{matrixele} are consistent with the matrix elements of the total wave functions constructed from the direct product between the flavor wave functions ($|\bm{8}^\prime,0,1\rangle$, $|\bm{10},0,1\rangle$) and the spin wave functions ($J=\frac{1}{2}$, $\frac{3}{2}$, $\frac{5}{2}$).

\begin{table*}[htbp]
\setlength\tabcolsep{1pt} \caption{The matrix elements $[\langle\mathcal{O}^{\text{f}}\rangle, \langle\mathcal{O}^{\text{fs}}\rangle]$ of the $\Sigma_c^{(*)}\bar{D}_s^{*}-\Xi_c^{\prime(*)} \bar{D}^{(*)}$ (left) and $\Xi_c^{\prime(*)}\bar{D}_s-\Omega_c^{(*)}\bar{D}^{(*)}$ (right) coupled channels. The eigenvalues $\langle\mathcal{O}^{\text{f}}_{\text{Eig}}\rangle$ and $\langle\mathcal{O}^{\text{fs}}_{\text{Eig}}\rangle$ are calculated in the SU(3) limit with $g_x=1.0$.\label{matrixele}}
\renewcommand\arraystretch{1.5}
\begin{tabular}{cccccccc}
\hline
&$[\langle\mathcal{O}^{\text{f}}\rangle,\langle\mathcal{O}^{\text{fs}}\rangle]$
&$\langle\mathcal{O}^{\text{f}}_{\text{Eig}}\rangle$&$\langle\mathcal{O}^{\text{fs}}_{\text{Eig}}\rangle$
&&$[\langle\mathcal{O}^{\text{f}}\rangle,\langle\mathcal{O}^{\text{fs}}\rangle]$
&$\langle\mathcal{O}^{\text{f}}_{\text{Eig}}\rangle$&$\langle\mathcal{O}^{\text{fs}}_{\text{Eig}}\rangle$\\
\hline
($|\Sigma_c\bar{D}_s\rangle^1_{\frac{1}{2}},$&\multirow{2}{*}{$\left(
\begin{array}{cc}
[-\frac{4}{3},0]&[2\sqrt{2}g_x,0] \\
& [\frac{2}{3},0]\\
\end{array}
\right)$}&\multirow{2}{*}{$\left(
\begin{array}{c}
-\frac{10}{3} \\
 \frac{8}{3}\\
\end{array}
\right)$}&\multirow{2}{*}{$\left(
\begin{array}{c}
0 \\
0 \\
\end{array}
\right)$}&($|\Xi^\prime_c\bar{D}_s\rangle^{\frac{1}{2}}_{\frac{1}{2}},$&\multirow{2}{*}{$\left(
\begin{array}{cc}
[\frac{2}{3},0]&[2\sqrt{2}g_x,0] \\
& [-\frac{4}{3},0]\\
\end{array}
\right)$}&\multirow{2}{*}{$\left(
\begin{array}{c}
-\frac{10}{3} \\
 \frac{8}{3}\\
\end{array}
\right)$}&\multirow{2}{*}{$\left(
\begin{array}{c}
 0\\
 0\\
\end{array}
\right)$}\\
$|\Xi^\prime_c\bar{D}\rangle^{1}_{\frac{1}{2}})$&&&&$|\Omega_c\bar{D}\rangle^{1}_{\frac{1}{2}})$\\

($|\Sigma^*_c\bar{D}_s\rangle^1_{\frac{3}{2}},$&\multirow{2}{*}{$\left(
\begin{array}{cc}
[-\frac{4}{3},0]& [2\sqrt{2}g_x,0]\\
& [\frac{2}{3},0]\\
\end{array}
\right)$}&\multirow{2}{*}{$\left(
\begin{array}{c}
 -\frac{10}{3}\\
 \frac{8}{3}\\
\end{array}
\right)$}&\multirow{2}{*}{$\left(
\begin{array}{c}
 0\\
 0\\
\end{array}
\right)$}&($|\Xi^*_c\bar{D}_s\rangle^{\frac{1}{2}}_{\frac{3}{2}},$&\multirow{2}{*}{$\left(
\begin{array}{cc}
[\frac{2}{3},0]& [2\sqrt{2}g_x,0]\\
& [-\frac{4}{3},0]\\
\end{array}
\right)$}&\multirow{2}{*}{$\left(
\begin{array}{c}
-\frac{10}{3} \\
 \frac{8}{3}\\
\end{array}
\right)$}&\multirow{2}{*}{$\left(
\begin{array}{c}
 0\\
 0\\
\end{array}
\right)$}\\
$|\Xi^*_c\bar{D}\rangle^{1}_{\frac{3}{2}})$&&&&$|\Omega^*_c\bar{D}\rangle^{\frac{1}{2}}_{\frac{3}{2}})$\\

($|\Sigma_c\bar{D}^*_s\rangle^1_{\frac{1}{2}},$&\multirow{2}{*}{$\left(
\begin{array}{cc}
[-\frac{4}{3},\frac{16}{9}]&[2\sqrt{2}g_x,-\frac{8\sqrt{2}}{3}] \\
&[\frac{2}{3},-\frac{8}{9}] \\
\end{array}
\right)$}&\multirow{2}{*}{$\left(
\begin{array}{c}
-\frac{10}{3} \\
 \frac{8}{3}\\
\end{array}
\right)$}&\multirow{2}{*}{$\left(
\begin{array}{c}
 \frac{40}{9}\\
 -\frac{32}{9}\\
\end{array}
\right)$}&($|\Xi^\prime_c\bar{D}^*_s\rangle^{\frac{1}{2}}_{\frac{1}{2}},$&\multirow{2}{*}{$\left(
\begin{array}{cc}
[\frac{2}{3},-\frac{8}{9}]&[2\sqrt{2}g_x,-\frac{8\sqrt{2}}{3}] \\
& [-\frac{4}{3},\frac{16}{9}]\\
\end{array}
\right)$}&\multirow{2}{*}{$\left(
\begin{array}{c}
-\frac{10}{3} \\
 \frac{8}{3}\\
\end{array}
\right)$}&\multirow{2}{*}{$\left(
\begin{array}{c}
 \frac{40}{9}\\
 -\frac{32}{9}\\
\end{array}
\right)$}\\
$|\Xi^\prime_c\bar{D}^*\rangle^{1}_{\frac{1}{2}})$&&&&$|\Omega_c\bar{D}^*\rangle^{\frac{1}{2}}_{\frac{1}{2}})$\\

($|\Sigma_c\bar{D}^*_s\rangle^1_{\frac{3}{2}},$&\multirow{2}{*}{$\left(
\begin{array}{cc}
[-\frac{4}{3},-\frac{8}{9}]& [2\sqrt{2}g_x,\frac{4\sqrt{2}}{3}]\\
& [\frac{2}{3},\frac{4}{9}]\\
\end{array}
\right)$}&\multirow{2}{*}{$\left(
\begin{array}{c}
-\frac{10}{3} \\
 \frac{8}{3}\\
\end{array}
\right)$}&\multirow{2}{*}{$\left(
\begin{array}{c}
 -\frac{20}{9}\\
 \frac{16}{9}\\
\end{array}
\right)$}&($|\Xi^\prime_c\bar{D}^*_s\rangle^{\frac{1}{2}}_{\frac{3}{2}},$&\multirow{2}{*}{$\left(
\begin{array}{cc}
[\frac{2}{3},\frac{4}{9}]& [2\sqrt{2}g_x,\frac{4\sqrt{2}}{3}]\\
& [-\frac{4}{3},-\frac{8}{9}]\\
\end{array}
\right)$}&\multirow{2}{*}{$\left(
\begin{array}{c}
 -\frac{10}{3}\\
 \frac{8}{3}\\
\end{array}
\right)$}&\multirow{2}{*}{$\left(
\begin{array}{c}
 -\frac{20}{9}\\
 \frac{16}{9}\\
\end{array}
\right)$}\\
$|\Xi^\prime_c\bar{D}^*\rangle^{1}_{\frac{3}{2}})$&&&&$|\Omega_c\bar{D}^*\rangle^{\frac{1}{2}}_{\frac{3}{2}})$\\

($|\Sigma^*_c\bar{D}^*_s\rangle^1_{\frac{1}{2}},$&\multirow{2}{*}{$\left(
\begin{array}{cc}
[-\frac{4}{3},\frac{20}{9}]& [2\sqrt{2}g_x,-\frac{10\sqrt{2}}{3}]\\
& [\frac{2}{3},-\frac{10}{9}]\\
\end{array}
\right)$}&\multirow{2}{*}{$\left(
\begin{array}{c}
-\frac{10}{3} \\
 \frac{8}{3}\\
\end{array}
\right)$}&\multirow{2}{*}{$\left(
\begin{array}{c}
 \frac{50}{9}\\
 -\frac{40}{9}\\
\end{array}
\right)$}&($|\Xi^*_c\bar{D}^*_s\rangle^{\frac{1}{2}}_{\frac{1}{2}},$&\multirow{2}{*}{$\left(
\begin{array}{cc}
[\frac{2}{3},-\frac{10}{9}]&[2\sqrt{2}g_x,-\frac{10\sqrt{2}}{3}] \\
& [-\frac{4}{3},\frac{20}{9}]\\
\end{array}
\right)$}&\multirow{2}{*}{$\left(
\begin{array}{c}
 -\frac{10}{3}\\
 \frac{8}{3}\\
\end{array}
\right)$}&\multirow{2}{*}{$\left(
\begin{array}{c}
 \frac{50}{9}\\
 -\frac{40}{9}\\
\end{array}
\right)$}\\
$|\Xi^*_c\bar{D}^*\rangle^{1}_{\frac{1}{2}})$&&&&$|\Omega^*_c\bar{D}^*\rangle^{\frac{1}{2}}_{\frac{1}{2}})$\\

($|\Sigma^*_c\bar{D}^*_s\rangle^1_{\frac{3}{2}},$&\multirow{2}{*}{$\left(
\begin{array}{cc}
[-\frac{4}{3},\frac{8}{9}]& [2\sqrt{2}g_x,-\frac{4\sqrt{2}}{3}]\\
& [\frac{2}{3},-\frac{4}{9}]\\
\end{array}
\right)$}&\multirow{2}{*}{$\left(
\begin{array}{c}
-\frac{10}{3} \\
 \frac{8}{3}\\
\end{array}
\right)$}&\multirow{2}{*}{$\left(
\begin{array}{c}
 \frac{20}{9}\\
 -\frac{16}{9}\\
\end{array}
\right)$}&($|\Xi^*_c\bar{D}^*_s\rangle^{\frac{1}{2}}_{\frac{3}{2}},$&\multirow{2}{*}{$\left(
\begin{array}{cc}
[\frac{2}{3},-\frac{4}{9}]& [2\sqrt{2}g_x,-\frac{4\sqrt{2}}{3}]\\
& [-\frac{4}{3},\frac{8}{9}]\\
\end{array}
\right)$}&\multirow{2}{*}{$\left(
\begin{array}{c}
-\frac{10}{3} \\
 \frac{8}{3}\\
\end{array}
\right)$}&\multirow{2}{*}{$\left(
\begin{array}{c}
 \frac{20}{9}\\
 -\frac{16}{9}\\
\end{array}
\right)$}\\
$|\Xi^*_c\bar{D}^*\rangle^{1}_{\frac{3}{2}})$&&&&$|\Omega^*_c\bar{D}^*\rangle^{\frac{1}{2}}_{\frac{3}{2}})$\\

($|\Sigma^*_c\bar{D}^*_s\rangle^1_{\frac{5}{2}},$&\multirow{2}{*}{$\left(
\begin{array}{cc}
[-\frac{4}{3},-\frac{4}{3}]&[2\sqrt{2}g_x,2\sqrt{2}] \\
& [\frac{2}{3},\frac{2}{3}]\\
\end{array}
\right)$}&\multirow{2}{*}{$\left(
\begin{array}{c}
-\frac{10}{3} \\
 \frac{8}{3}\\
\end{array}
\right)$}&\multirow{2}{*}{$\left(
\begin{array}{c}
 -\frac{10}{3}\\
 \frac{8}{3}\\
\end{array}
\right)$}&($|\Xi^*_c\bar{D}^*_s\rangle^{\frac{1}{2}}_{\frac{5}{2}},$&\multirow{2}{*}{$\left(
\begin{array}{cc}
[\frac{2}{3},\frac{2}{3}]&[2\sqrt{2}g_x,2\sqrt{2}] \\
& [-\frac{4}{3},-\frac{4}{3}]\\
\end{array}
\right)$}&\multirow{2}{*}{$\left(
\begin{array}{c}
 -\frac{10}{3}\\
 \frac{8}{3}\\
\end{array}
\right)$}&\multirow{2}{*}{$\left(
\begin{array}{c}
 -\frac{10}{3}\\
 \frac{8}{3}\\
\end{array}
\right)$}\\
$|\Xi^*_c\bar{D}^*\rangle^{1}_{\frac{5}{2}})$&&&&$|\Omega^*_c\bar{D}^*\rangle^{\frac{1}{2}}_{\frac{5}{2}})$\\
\hline
\end{tabular}
\end{table*}

To find the possible bound states from the coupled channel effective potentials, we solve the two-channel Lippmann-Schwinger equation (LSE)
\begin{eqnarray}
\mathbb{T}(E)=\mathbb{V}+\mathbb{V}\mathbb{G}\mathbb{T}(E),
\end{eqnarray}
with
\begin{eqnarray}
\mathbb{V}=\left(
             \begin{array}{cc}
               v_{11} & v_{12} \\
               v_{21} & v_{22} \\
             \end{array}
           \right),
\end{eqnarray}
and
\begin{eqnarray}
\mathbb{T}=\left(
             \begin{array}{cc}
               t_{11} & t_{12} \\
               t_{21} & t_{22} \\
             \end{array}
           \right).
\end{eqnarray}
The $2\times2$ matrix $\mathbb{G}$ is defined as
\begin{eqnarray}
\mathbb{G}(E)=\left(
             \begin{array}{cc}
               G_{1} & 0 \\
               0 & G_{2} \\
             \end{array}
           \right),
\end{eqnarray}
with
\begin{eqnarray}
G_i=\frac{1}{2\pi^2}\int dq\frac{q^2}{E-\sqrt{m_{i1}^2+q^2}-\sqrt{m_{i2}^2+q^2}}u^{2}(\Lambda).\nonumber\\
\end{eqnarray}
The $m_{i1}$ and $m_{i2}$ are the masses of the single-charm baryon and anticharmed meson in the $i$-th channels with $i=$1, 2.
We introduce a dipole form factor \cite{Nakamura:2022gtu,Chen:2022wkh,Leinweber:2003dg,Wang:2007iw}
\begin{eqnarray}
u(\Lambda)=\frac{1}{1+q^2/\Lambda^2}
\end{eqnarray}
to suppress the contributions from higher momenta, the regular parameter $\Lambda$ is fixed at 1.0 GeV, this cutoff together with the parameters in scenario 1 or scenario 2 have been adopted to successfully reproduce the masses of the observed $P_c$, $P_{cs}$ \cite{Chen:2024tuu}, and $T_{cc}$ \cite{Chen:2024bre} states.

The pole position of LSE satisfies $|\bm{1}-\mathbb{V}\mathbb{G}|=0$. For the possible bound states below the lowest channel, we search the bound state solutions in the first Riemann sheet of the lowest channel. For the possible quasi-bound states between the thresholds of the first and second channels, we replace the integration variable
\begin{eqnarray}
q\rightarrow q\times\text{exp}(-i\theta),\label{theta}
\end{eqnarray}
with $0<\theta<\frac{\pi}{2}$ to find the solutions in the second Riemann sheet of the lower channel and first Riemann of the higher channel.

\begin{figure*}[!htbp]
    \centering
    \includegraphics[width=0.8\linewidth]{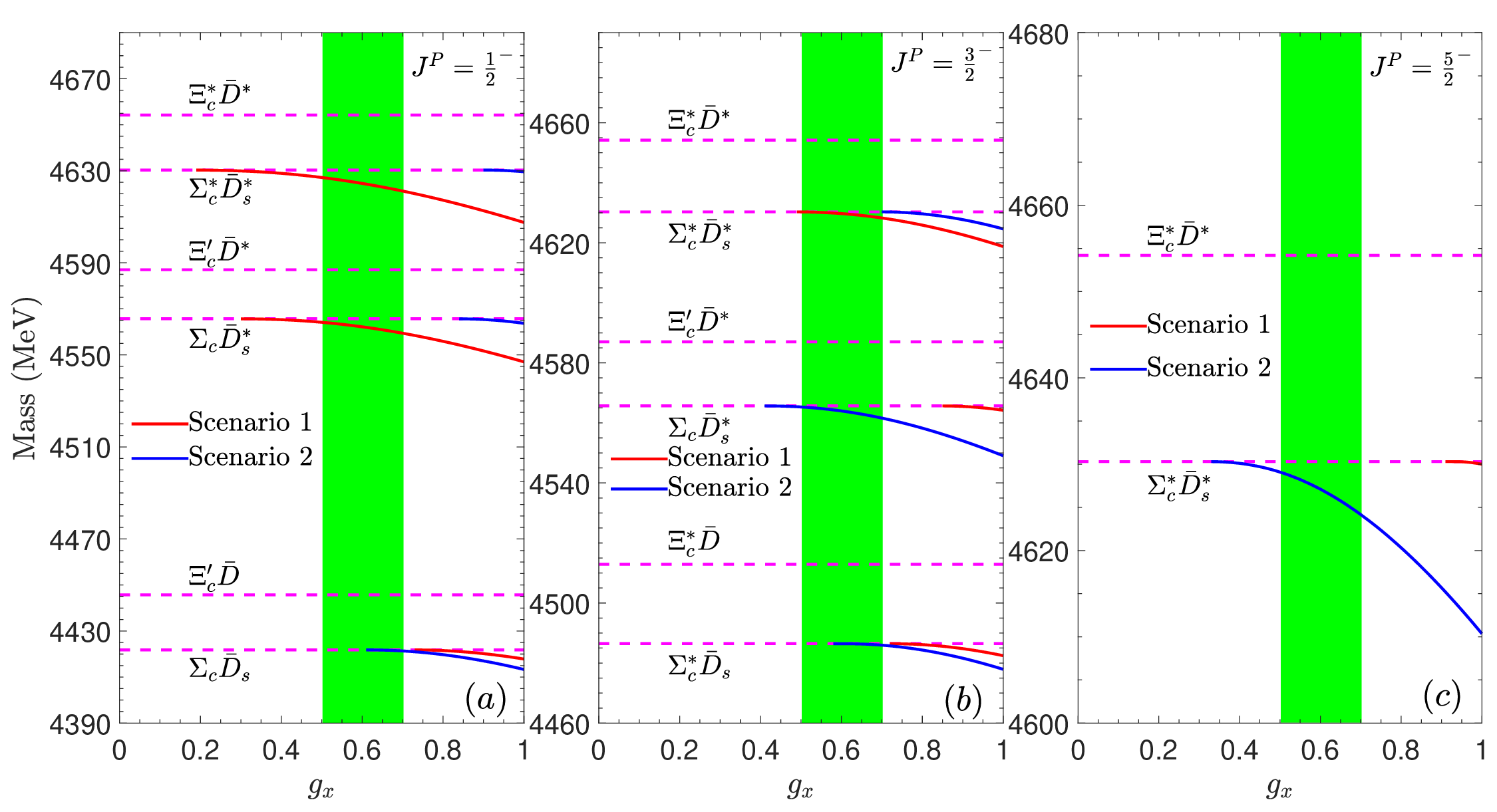}
    \caption{The $g_x$ dependences of the pole positions produced from the $\Sigma_c^{*}\bar{D}^{(*)}_s$-$\Xi_c^{\prime(*)}\bar{D}^{(*)}$ mixing with $J^P=\frac{1}{2}^-$ (a), $\frac{3}{2}^-$ (b), and $\frac{5}{2}^-$ (c). The pole trajectories obtained from scenario 1 and scenario 2 are illustrated with red and blue lines, respectively. The physical region with $g_x=0.6\pm0.1$ are illustrated with green bands.}
    \label{Pcs1}
\end{figure*}

\begin{figure*}[!htbp]
    \centering
    \includegraphics[width=0.8\linewidth]{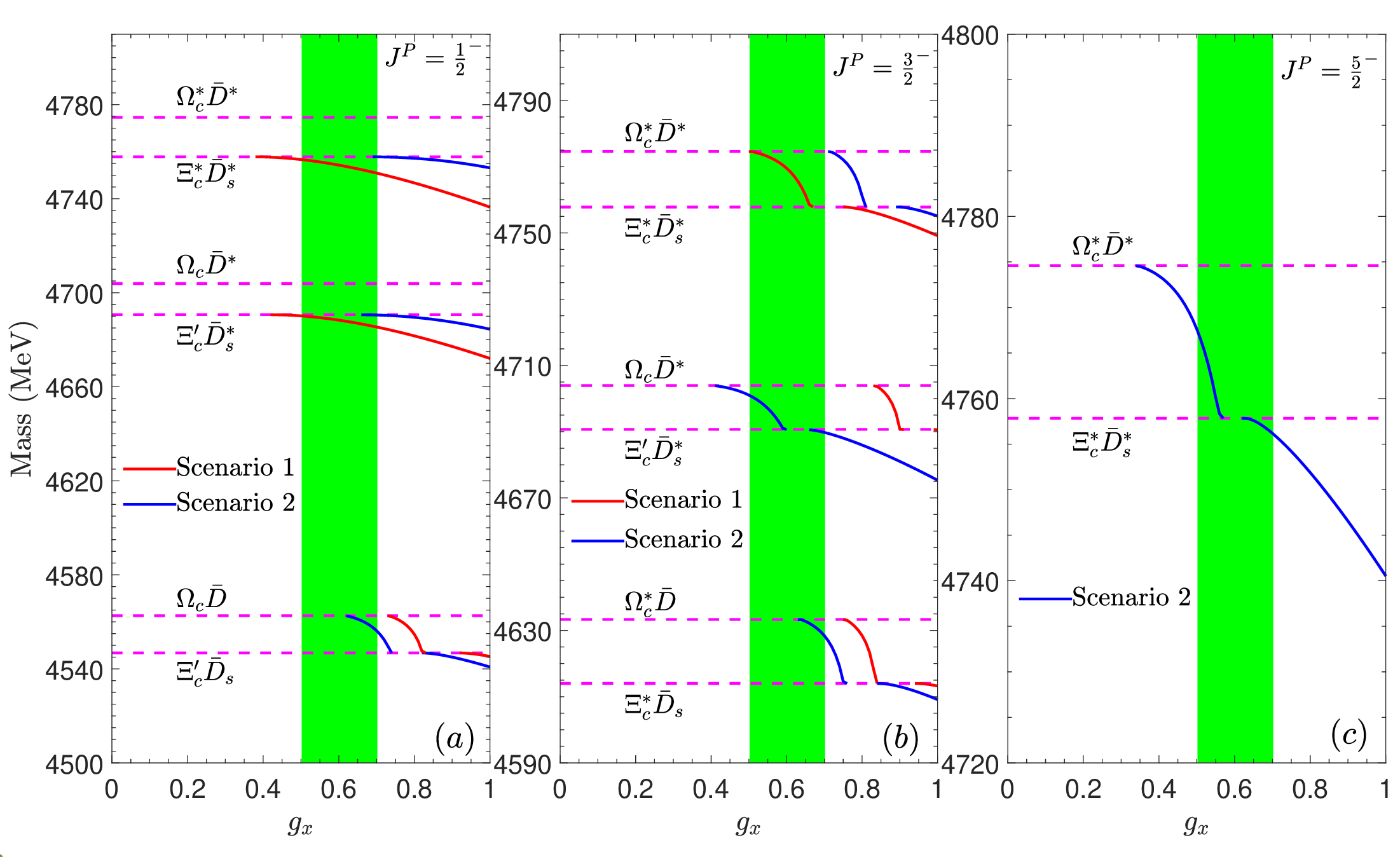}
    \caption{The $g_x$ dependences of the pole positions produced from the $\Xi_c^{\prime(*)}\bar{D}_s^{(*)}$-$\Omega_c^{(*)}\bar{D}^{(*)}$ mixing with $J^P=\frac{1}{2}^-$ (a), $\frac{3}{2}^-$ (b), and $\frac{5}{2}^-$ (c). The pole trajectories obtained from scenario 1 and scenario 2 are illustrated with red and blue lines, respectively. The physical region with $g_x=0.6\pm0.1$ are illustrated with green bands.}
    \label{Pcss}
\end{figure*}


In Fig. \ref{Pcs1}, the $g_x$-dependences of the pole trajectories for the single-strange hidden-charm pentaquarks obtained from scenario 1 and scenario 2 are illustrated with red and blue lines, respectively. The results with $J^P=\frac{1}{2}^-$, $\frac{3}{2}^-$, and $J=\frac{5}{2}^-$ are illustrated in Figs. \ref{Pcs1} (a), (b), and (c), respectively. We use the mass of $P_{cs}(4338)$ to fix the $g_x$ at interval $0.6\pm0.1$, we illustrate this physical region with a green band.

As shown in Fig. \ref{Octth}, the threshold of $\Sigma^{(*)}_c\bar{D}_s^{(*)}$ is lower than the threshold of $\Xi_c^{\prime(*)}\bar{D}^{(*)}$ by about 20-30 MeV. From Table \ref{matrixele}, we find that the diagonal effective potential $V^{\Sigma_c^{(*)}\bar{D}_s^{(*)}\rightarrow \Sigma_c^{(*)}\bar{D}_s^{(*)}}$ from the lower channel is weakly attractive, while the diagonal effective potential $V^{\Xi_c^{\prime(*)}\bar{D}^{(*)}\rightarrow \Xi_c^{\prime(*)}\bar{D}^{(*)}}$ from the higher channel is repulsive.
After the inclusion of the off-diagonal $V^{\Sigma_c^{(*)}\bar{D}_s^{(*)}\rightarrow \Xi_c^{\prime(*)}\bar{D}^{(*)}}$ ($V^{\Xi_c^{\prime(*)}\bar{D}^{(*)}\rightarrow \Sigma_c^{(*)}\bar{D}_s^{(*)}}$) potential, as illustrated in Fig. \ref{Pcs1}, in both scenarios, by varying the $g_x$ in the [0,1] region, the bound state poles can only appear below the lower $\Sigma_c^{(*)}\bar{D}_s^{(*)}$ channel.

In Fig. \ref{Pcs1}, we also show how the parameters under scenario 1 and scenario 2 affect the formations of molecular states with different $J$. As shown in Fig. \ref{Pcs1} (a), in scenario 1, at $0.73\leq g_x\leq 1$, the $\Sigma_c\bar{D}_s-\Xi_c^\prime\bar{D}$ mixing leads to a bound state with its mass below the lower $\Sigma_c\bar{D}_s$ channel, while in scenario 2, at $0.61\leq g_x\leq 1$, the $\Sigma_c\bar{D}_s-\Xi_c^\prime\bar{D}$ mixing  can produce a bound state, its pole trajectory overlaps with the physical $g_x$ region.

For the $J^{P}=\frac{3}{2}^-$ poles illustrated in Fig. \ref{Pcs1} (b), we find that in scenario 1, only the pole trajectory below the $\Sigma_c^*\bar{D}_s^*$ threshold overlaps with the green physical region, and in scenario 2, the pole trajectories below the $\Sigma_c^*\bar{D}_s$ and $\Sigma_c\bar{D}_s^*$ thresholds overlap with the green physical region and thus are possible to form bound states.

Finally, for the $J^{P}=\frac{5}{2}^-$ states, we find that only the pole trajectory obtained from scenario 2 overlaps with green physical region, as illustrated in Fig. \ref{Pcs1} (c).

In Fig. \ref{Pcss}, the $g_x$-dependences of the pole trajectories for the double-strange hidden-charm pentaquarks obtained under scenario 1 and scenario 2 are illustrated with red and blue lines, respectively. By checking the thresholds of the relevant $\Xi_c^{\prime(*)}\bar{D}_s^{(*)}$ and $\Omega_c^{(*)}\bar{D}^{(*)}$ channels in Fig. \ref{Octth}, we find that the threshold of the $\Xi_c^{(\prime)*}\bar{D}_s^{(*)}$ is lower than the threshold of $\Omega_c^{(*)}\bar{D}^{(*)}$ by about 10-20 MeV. As shown in Table \ref{matrixele}, the diagonal effective potential $V^{\Xi_c^{\prime(*)}\bar{D}_s^{(*)}\rightarrow\Xi_c^{\prime(*)}\bar{D}_s^{(*)}}$ from the lower channel is weakly repulsive, while the diagonal effective potential $V^{\Omega_c^{(*)}\bar{D}^{(*)}\rightarrow\Omega_c^{(*)}\bar{D}^{(*)}}$ from the higher channel is weakly attractive, which is different from the single-strange hidden-charm pentaquark case.

Comparing to the $\Sigma_c^{*}\bar{D}^{(*)}_s$-$\Xi_c^{\prime(*)}\bar{D}^{(*)}$ mixing case, the properties of the interactions of the higher channel and the lower channel is reversed in the $\Xi_c^{\prime(*)}\bar{D}_s^{(*)}$-$\Omega_c^{(*)}\bar{D}^{(*)}$ mixing case. Consequently, as shown in Fig. \ref{Pcss}, the poles produced via $\Xi_c^{\prime(*)}\bar{D}_s^{(*)}$-$\Omega_c^{(*)}\bar{D}^{(*)}$ mixing may appear between the thresholds of two channels, or below the channel with the lower threshold.

Explicitly, for the $J^P=\frac{1}{2}^-$ states, we find that the pole trajectories below the $\Xi_c^{\prime}\bar{D}_s^{*}$ and $\Xi_c^*\bar{D}_s^*$ overlap with the green physical region in scenario 1, while in scenario 2, the overlaps between the pole trajectories of these two states and the green physical are very small. For the pole close to the $\Xi^\prime\bar{D}_s$/$\Omega_c\bar{D}$ threshold, we find that in scenario 2 the blue pole trajectory has a small overlap with the green physical region, i.e., when the $0.61<g_x<0.75$, besides, in this scenario, when $g_x$ further increases to 0.82, at $0.75<g_x<0.82$, this region is expected to exist a virtual state at the negative imaginary momentum plane, this pole might be found with an improved complex scaling method \cite{Chen:2023eri}. In our present work, we only focus on the bound and resonance states by varying the $\theta$ defined in Eq. (\ref{theta}). When the $g_x$ further increases in the region $0.82<g_x<1$, this pole becomes a bound state.

For the $J^P=\frac{3}{2}^-$ poles, their behaviors can be discussed in a similar way, and for the $J^P=\frac{5}{2}^-$, we do not find any poles in scenario 1, thus we only illustrate the pole trajectory calculated from scenario 2.

In Table \ref{8results}, we further present our numerical results for the possible bound/quasi-bound states produced from the $\Sigma_c^{*}\bar{D}^{(*)}_s$-$\Xi_c^{\prime(*)}\bar{D}^{(*)}$ and $\Xi_c^{\prime(*)}\bar{D}_s^{(*)}$-$\Omega_c^{(*)}\bar{D}^{(*)}$ mixings. Since in Ref. \cite{Chen:2024tuu} we have already calculated the possible bound/quasi bound states with flavors $|\bm{8}^\prime,1,\frac{1}{2}\rangle$ ($\Sigma_c^{(*)}\bar{D}^{(*)}$ $I=\frac{1}{2}$) and $|\bm{8}^\prime,0,0\rangle$ ($\Xi_c^{\prime(*)}\bar{D}^{(*)}$ $I=0$) in both scenarios, we also present these results in Table \ref{8results} for comparison.

From the results listed in Table \ref{8results}, we find that in both scenarios, three $J^{P}=\frac{1}{2}^{-}$, three $J^P=\frac{3}{2}^-$, and one $J^{P}=\frac{5}{2}^-$ can be find for the states with flavors $|\bm{8}^\prime,1,\frac{1}{2}\rangle$ and $|\bm{8}^\prime,0,0\rangle$. However, the SU(3) flavor wave functions $|\bm{8}^\prime,0,1\rangle$ and $|\bm{8}^\prime,-1,\frac{1}{2}\rangle$ will reduce to the $\Sigma^{(*)}_c\bar{D}^{(*)}_s-\Xi_c^{\prime(*)}\bar{D}^{(*)}$ ($I=1$) and $\Xi_c^{\prime(*)}\bar{D}_s^{(*)}-\Omega_c^{(*)}\bar{D}^{(*)}$ ($I=\frac{1}{2}$) states via SU(3) breaking, thus some of the states can not gain enough attractive forces to form bound/quasi-bound states.

\begin{table*}[htbp]
\renewcommand\arraystretch{1.5}
\caption{The masses of the bound/quasi-bound states with flavors $\Sigma^{(*)}_c\bar{D}^{(*)}_s-\Xi_c^{\prime(*)}\bar{D}^{(*)}$ and $\Xi_c^{\prime(*)}\bar{D}_s^{(*)}-\Omega_c^{(*)}\bar{D}^{(*)}$. The masses of the bound/quasi-bound states with flavors ($I=\frac{1}{2}$) $\Sigma_c^{(*)}\bar{D}^{(*)}$ and ($I=0$) $\Xi_c^{\prime(*)}\bar{D}^{(*)}$ obtained in Ref. \cite{Chen:2024tuu} are also presented for comparison. The "$\dagger$" denotes the state we take as input. These results are obtained by considering both scenarios.}
\begin{tabular}{c|ccc|c|cccccccccc}
\toprule[0.8pt]
Components&$I(J^P)$&Scenario 1&Scenario 2&Components&$I(J^P)$&Scenario 1&Scenario 2\\
\hline
$\Sigma_c\bar{D}_s-\Xi_c^\prime\bar{D}$&$1(\frac{1}{2}^-)$&-&$4421.8^{+*}_{-0.4}$
&$\Xi_c^\prime\bar{D}_s-\Omega_c\bar{D}$&$\frac{1}{2}(\frac{1}{2}^-)$&-&$4562.6^{+*}_{-6.3}$\\
$\Sigma_c\bar{D}^*_s-\Xi_c^\prime\bar{D}^*$&$1(\frac{1}{2}^-)$&$4562.2^{+1.8}_{-2.8}$&-
&$\Xi_c^\prime\bar{D}^*_s-\Omega_c\bar{D}^*$&$\frac{1}{2}(\frac{1}{2}^-)$&$4688.4^{+1.8}_{-3.0}$&-\\
$\Sigma^*_c\bar{D}^*_s-\Xi_c^*\bar{D}^*$&$1(\frac{1}{2}^-)$&$4624.5^{+2.6}_{-3.3}$&-
&$\Xi_c^\prime\bar{D}^*_s-\Omega^*_c\bar{D}^*$&$\frac{1}{2}(\frac{1}{2}^-)$&$4754.3^{+2.4}_{-3.4}$&-\\
\hline
$\Sigma^*_c\bar{D}_s-\Xi_c^*\bar{D}_s$&$1(\frac{3}{2}^-)$&-&$4486.5^{+*}_{-0.5}$
&$\Xi_c^*\bar{D}_s-\Omega^*_c\bar{D}$&$\frac{1}{2}(\frac{3}{2}^-)$&-&$4633.3^{+*}_{-4.9}$\\
$\Sigma_c\bar{D}^*_s-\Xi_c^\prime\bar{D}^*$&$1(\frac{3}{2}^-)$&-&$4564.0^{+1.3}_{-2.4}$
&$\Xi_c^\prime\bar{D}^*_s-\Omega_c\bar{D}^*$&$\frac{1}{2}(\frac{3}{2}^-)$&-&$4690.7^{+10.3}_{-0.9}$\\
$\Sigma^*_c\bar{D}^*_s-\Xi_c^*\bar{D}^*$&$1(\frac{3}{2}^-)$&$4629.7^{+0.6}_{-1.4}$&-
&$\Xi_c^*\bar{D}^*_s-\Omega^*_c\bar{D}^*$&$\frac{1}{2}(\frac{3}{2}^-)$&$4769.5^{+5.1}_{-11.7}$&-\\
\hline
$\Sigma^*_c\bar{D}^*_s-\Xi_c^*\bar{D}^*$&$1(\frac{5}{2}^-)$&-&$4627.1^{+2.0}_{-2.9}$
&$\Xi_c^*\bar{D}^*_s-\Omega^*_c\bar{D}^*$&$\frac{1}{2}(\frac{5}{2}^-)$&-&$4757.8_{-1.6}^{+9.8}$\\
\hline
Components&$I(J^P)$&Scenario 1&Scenario 2&Components&$I(J^P)$&Scenario 1&Scenario 2\\
\hline
$\Sigma_c\bar{D}$&$\frac{1}{2}(\frac{1}{2}^-)$&$4308.2_{-13.2}^{+2.2}$&$4305.3_{-4.2}^{+3.9}$
&$\Xi_c^\prime\bar{D}$&$0(\frac{1}{2}^-)$&$4433.8_{-4.9}^{+3.4}$&$4430.0_{-4.3}^{+4.1}$\\
$\Sigma_c\bar{D}^*$&$\frac{1}{2}(\frac{1}{2}^-)$&${}^{\dagger}4440.3_{-5.0}^{+4.0}$&${}^{\dagger}4457.3_{-1.8}^{+4.0}$
&$\Xi_c^\prime\bar{D}^*$&$0(\frac{1}{2}^-)$&$4565.4_{-4.5}^{+4.6}$&$4582.5_{-2.2}^{+4.6}$\\
$\Sigma^*_c\bar{D}^*$&$\frac{1}{2}(\frac{1}{2}^-)$&$4501.4_{-6.2}^{+5.0}$&$4523.8_{-2.4}^{+2.8}$
&$\Xi_c^*\bar{D}^*$&$0(\frac{1}{2}^-)$&$4628.5_{-5.5}^{+4.8}$&$4651.4_{-2.6}^{+3.2}$\\
\hline
$\Sigma^*_c\bar{D}$&$\frac{1}{2}(\frac{3}{2}^-)$&$4373.3_{-6.8}^{+3.4}$&$4369.6_{-4.3}^{+4.0}$
&$\Xi_c^*\bar{D}$&$0(\frac{3}{2}^-)$&$4501.7_{-3.5}^{+4.5}$&$4496.9_{-4.2}^{+4.2}$\\
$\Sigma_c\bar{D}^*$&$\frac{1}{2}(\frac{3}{2}^-)$&${}^{\dagger}4457.3_{-1.8}^{+4.0}$&${}^{\dagger}4440.3_{-5.0}^{+4.0}$
&$\Xi_c^\prime\bar{D}^*$&$0(\frac{3}{2}^-)$&$4582.1_{-2.0}^{+4.2}$&$4564.5_{-5.1}^{+4.1}$\\
$\Sigma^*_c\bar{D}^*$&$\frac{1}{2}(\frac{3}{2}^-)$&$4513.4_{-3.2}^{+5.7}$&$4518.0_{-2.5}^{+4.9}$
&$\Xi_c^*\bar{D}^*$&$0(\frac{3}{2}^-)$&$4640.7_{-3.1}^{+5.8}$&$4645.7_{-2.3}^{+5.0}$\\
\hline
$\Sigma^*_c\bar{D}^*$&$\frac{1}{2}(\frac{5}{2}^-)$&$4523.8_{-1.2}^{+2.0}$&$4501.6_{-2.2}^{+3.4}$
&$\Xi_c^*\bar{D}^*$&$0(\frac{5}{2}^-)$&$4651.2_{-1.3}^{+2.0}$&$4628.2_{-2.2}^{+3.3}$\\
\hline
\end{tabular}\label{8results}
\end{table*}

\section{summary}\label{sec6}
In this work, we construct flavor wave functions built from single-charm baryons ($\Lambda_c$, $\Xi_c$, $\Sigma_c^{(*)}$, $\Xi_c^{\prime(*)}$, $\Omega_c^{(*)}$) and anticharmed mesons ($\bar{D}^{(*)}$, $\bar{D}_s^{(*)}$) within both the SU(3) and isospin bases. We then derive the mapping relations between the flavor wave functions formulated in these two bases.

We describe the baryon-meson interaction by introducing the $\tilde{g}_s\bm{\lambda}_1\cdot\bm{\lambda}_2$ and $\tilde{g}_a\bm{\lambda}_1\cdot\bm{\lambda}_2\bm{\sigma}_1\cdot\bm{\sigma}_2$ contact terms. From our calculations of the $P_c$ and $P_{cs}$ states in Ref. \cite{Chen:2022wkh}, we suggest that the interactions between the single-charm baryons and anti-charm mesons may follow a flavor dominant selection rule. Under this rule, the states that belong to the $\bm{8}^\prime$ and $\bm{1}$ representations have attractive forces, while the states that belong to the $\bm{8}$ and $\bm{10}$ representations have repulsive forces.

Among the attractive $\bm{8}^\prime$ and $\bm{1}$ representations, the SU(3) flavor eigenstates $|\bm{8}^\prime,1,\frac{1}{2}\rangle$ and $|\bm{8}^\prime,0,0\rangle$ directly match the physical channels $\Sigma_c^{(*)}\bar{D}^{(*)}$ ($I=\frac{1}{2}$) and $\Xi_c^{\prime(*)}\bar{D}^{(*)}$ ($I=0$). Accordingly, the binding energies of the $\Sigma_c^{(*)}\bar{D}^{(*)}$ ($I=\frac{1}{2}$) systems are very similar to that of the $\Xi_c^{\prime(*)}\bar{D}^{(*)}$ ($I=0$) systems, manifesting a good SU(3) symmetry.

In contrast, the SU(3) eigenstates $|\bm{1},0,0\rangle$, $|\bm{8}^\prime,0,1\rangle$, and $|\bm{8}^{\prime},-1,\frac{1}{2}\rangle$ cannot be identified with a single physical meson-baryon channel. Owing to SU(3) flavor symmetry breaking, each pure SU(3) state decomposes into a linear admixture of two coupled channels, i.e., $\Lambda_c\bar{D}_s^{(*)}-\Xi_c\bar{D}^{(*)}$, $\Sigma_c^{(*)}\bar{D}_s^{(*)}-\Xi_c^{\prime(*)}\bar{D}^{(*)}$, and $\Xi_c^{\prime(*)}\bar{D}_s^{(*)}-\Omega_c^{(*)}\bar{D}^{(*)}$, respectively.

We discuss the SU(3) breaking effects of the $|\bm{1},0,0\rangle$, $|\bm{8}^\prime,0,1\rangle$, and $|\bm{8}^{\prime},-1,\frac{1}{2}\rangle$ states by adjusting the $g_x$ values in the interval $[0,1]$. We show that the mixtures of $\Lambda_c\bar{D}^{(*)}-\Xi_c\bar{D}^{(*)}$, $\Sigma_c^{(*)}\bar{D}_s^{(*)}-\Xi_c^{\prime(*)}\bar{D}^{(*)}$, and $\Xi_c^{\prime(*)}\bar{D}_s^{(*)}-\Omega_c^{(*)}\bar{D}^{(*)}$ yield distinct pole trajectories, depending on the effective potentials of each system.

We also present the masses of the $\Sigma_c^{(*)}\bar{D}^{(*)}$ ($I=\frac{1}{2}$), $\Xi_c^{\prime(*)}\bar{D}^{(*)}$ ($I=0$), $\Sigma_c^{(*)}\bar{D}_s^{(*)}-\Xi_c^{\prime(*)}\bar{D}^{(*)}$ ($I=1$), and $\Xi_c^{\prime(*)}\bar{D}_s^{(*)}-\Omega_c^{(*)}\bar{D}^{(*)}$ ($I=\frac{1}{2}$) estimated in both scenarios. We are looking forward to the upcoming experimental measurements and hope that our predictions can be tested in the future.
\section*{Acknowledgments}
This work is supported by the National Natural Science Foundation of China under Grants No. 12305090, and the Natural Science Foundation of Hebei Province under Grants No. A2025201016. K. Chen is also supported by the Start-up Funds of Northwest University.

\end{document}